\documentclass[aps,floats]{revtex4}
\usepackage{amsmath,amssymb}
\usepackage{graphicx,epsfig}

\begin{document}
\bibliographystyle {plain}

\def\oppropto{\mathop{\propto}} 
\def\opsimeq{\mathop{\simeq}}
\def\opoverderline{\mathop{\overline}}
\def\operarrow{\mathop{\longrightarrow}}
\def\opsim{\mathop{\sim}}

\def\fig#1#2{\includegraphics[height=#1]{#2}}
\def\figx#1#2{\includegraphics[width=#1]{#2}}


\title{ Disorder-dominated phases of random systems   : \\
relations between tails exponents and scaling exponents  } 


 \author{ C\'ecile Monthus and Thomas Garel }
  \affiliation{Service de Physique Th\'{e}orique, CEA/DSM/SPhT\\
 Unit\'e de recherche associ\'ee au CNRS\\
 91191 Gif-sur-Yvette cedex, France}

\begin{abstract}

We consider various random models
(directed polymer, ferromagnetic random Potts model, Ising spin-glasses) 
in their disorder-dominated phases, where the free-energy cost $F(L)$ of
an excitation of length $L$ present fluctuations 
that grow as a power-law $\Delta F(L) \sim L^{\omega}$
with the so-called droplet exponent $\omega>0$.
We study the tails of the probability distribution $\Pi(x)$
of the rescaled free-energy cost 
$x= \frac{F_L- \overline{F_L}}{L^{\omega}}$,
which are governed by two exponents $(\eta_-,\eta_+)$ defined by
 $\ln \Pi(x \to \pm \infty) \sim - \vert x \vert^{\eta_{\pm}}$. 
The aim of this paper is to establish simple relations between
these tail exponents $(\eta_-,\eta_+) $ and the droplet exponent $\omega$.
We first prove these relations for disordered models on diamond hierarchical 
lattices where exact renormalizations exist for the probability 
distribution $\Pi(x)$. We then interpret these relations
via an analysis of the measure of the
rare disorder configurations governing the tails.
Our conclusion is that these relations, when expressed
in terms of the dimensions of the bulk and of the excitation surface
 are actually valid for general lattices.

\end{abstract}

\maketitle

\section{ Introduction }

To understand the low-temperature phase of disordered systems,
it is important to characterize both
 the statistics of the ground state energy over the disordered samples,
and the statistics of excitations above the ground state
within one sample.

The ground-state energy $E_0$ of a disordered sample
is the minimal energy among the energies of all possible configurations.
The study of its distribution thus belongs to the field of extreme value statistics.
Whereas the case of independent random variables is well 
classified in three universality classes \cite{Gum_Gal}, the problem for 
the correlated energies within a disordered sample remains open
and has been the subject of many recent studies. 
The interest lies both \\
(i) in the scaling behavior of the average
 $E_0^{av}(L)$ and the standard deviation $ \Delta E_0(L)$ 
 with the linear size $L$ \\
(ii) in the asymptotic distribution $P(x)$ 
of the rescaled variable $x=(E_0 -E_0^{av}(L))/\Delta E_0(L)$
in the limit $L \to \infty$
\begin{equation}
{\cal P}_L(E_0)   \opsimeq_{L \to \infty}  \frac{1}{ \Delta E_0(L)} \  
\Pi \left( x= \frac{
E_0 -E_0^{av}(L)}{ \Delta E_0(L) }  \right) 
\label{scalinge0}
\end{equation}

For spin-glasses in finite dimension $d$, a sample of linear size $L$ 
contains $N=L^d$ spins. Following the definitions of Ref. \cite{Bou_Krz_Mar},
the `shift exponent' $\omega_s$ governs the correction to extensivity
of the averaged value
\begin{eqnarray}
E_0^{av}(L) \sim L^d e_0+ L^{\omega_s} e_1 +...
\label{e0av}
\end{eqnarray}
Within the droplet theory \cite{Fis_Hus_SG,Fis_Hus},
this shift exponent $\omega_s$ coincides
with the domain wall exponent $\omega_{DW}$ and
with the droplet exponent $\omega$ of low energy excitations (see below)
\begin{eqnarray}
\omega_s = \omega_{DW} = \omega
\label{omegashift}
\end{eqnarray}
The `fluctuation exponent' $\omega_f $  governs the growth
 of the standard deviation
 \begin{eqnarray}
\Delta E_0(L)  \sim  L^{\omega_f} 
\label{deltae0}
\end{eqnarray}
In any finite dimension $d$, it has been proven
that the fluctuation
exponent is $\omega_f=d/2$ \cite{We_Ai}.
Accordingly, the rescaled distribution $P(x)$ of Eq. (\ref{scalinge0}) 
was numerically found to be Gaussian in $d=2$ and $d=3$ 
\cite{Bou_Krz_Mar},
suggesting some Central Limit theorem.
In contrast with finite-dimensional
spin-glasses where one needs to introduce two exponents $\omega_s$
 and $\omega_f$,
the directed polymer model \cite{Hal_Zha} is characterized by
 a single exponent $\omega$ that governs both
 the correction to extensivity
of the average $E_0^{av}(L)$ and the width $\Delta E_0(L)$ 
\begin{eqnarray}
E_0^{av}(L) && \sim L e_0 + L^{\omega} e_1 +... \nonumber \\
\Delta E_0(L) && \sim  L^{\omega} e_2 +...
\end{eqnarray}
This exponent also governs the statistics of low excitations
within the droplet theory \cite{Fis_Hus}, as confirmed numerically
\cite{DPexcita}.
In dimension $1+1$, this exponent
is exactly known to be $\omega(d=1)=1/3$
\cite{Hus_Hen_Fis,Kar,Joh,Pra_Spo}
and the corresponding rescaled distribution $\Pi(x)$ 
 is related to Tracy-Widom distributions
of the largest eigenvalue of random matrices ensembles
 \cite{Joh,Pra_Spo,prae}.

Among disordered systems, the directed polymer model
thus presents the following distinctive  feature :
the statistics of the ground state energy over the samples
is directly related to the statistics of excitations within a given sample
\cite{Fis_Hus}. 
In finite-dimensional spin systems however,
the statistics of the ground-state energy over the samples
is not very interesting : the fluctuation exponent $\theta_f=d/2$
and the corresponding Gaussian distribution $P(x)$ simply reflects
the fluctuations of the $L^d$ random couplings defining the samples.
The only information it contains on the statistics of excitations
is the shift exponent $\omega_s=\omega$ governing the correction
to extensivity of the averaged value.
In this paper, we will be interested into the statistics 
of the energy $E_L$ of excitations above the ground state
in a given sample.
Its fluctuations are governed by the droplet exponent $\omega$
\begin{eqnarray}
\Delta E_L \sim L^{\omega}
\end{eqnarray}
and its distribution is expected to follow a 
scaling form as $L \to \infty$ 
\begin{equation}
P_L(E_L)   \opsimeq_{L \to \infty}  \frac{1}{ L^{\omega}} \  
{\widetilde \Pi} \left( x= \frac{
E_L - \overline{E_L }}{ L^{\omega} }  \right) 
\label{scalingexci}
\end{equation}
Our aim is to show that the exponents governing the tails
of the rescaled distributions ${\widetilde \Pi}(x)$ 
are simply related to the droplet exponent $\omega$.
Since the whole low-temperature phase ($T <T_c$) is 
described at large scale by the zero-temperature fixed point,
the statistics of the free-energy $F_L$ of excitations at $T<T_c$
is the same as the statistics of the energy $E_L$ of excitations 
above the ground state, up to some rescaling with the correlation length 
$\xi(T)$.
So the results concerning the tails of ${\widetilde \Pi}(x)$ 
also concerns the tails
of the free-energy of excitations at any temperature $T<T_c$.

\begin{figure}[htbp]
\includegraphics[height=6cm]{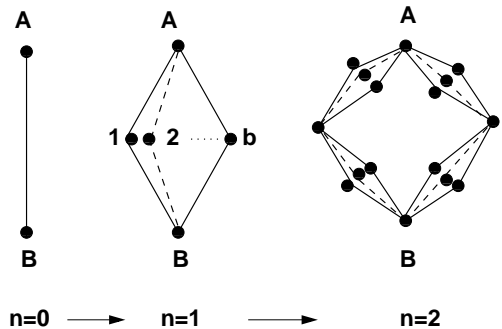}
\hspace{1cm}
\caption{ Hierarchical construction of the diamond lattice of
branching ratio $b$.   }
\label{figdiamond}
\end{figure}

To establish the relations existing between the tails of the probability
distribution ${\widetilde \Pi}(x)$ and the droplet exponent $\omega$,
we will first focus on the diamond hierarchical 
lattices where exact renormalizations exist as we now recall.
Among real-space renormalization procedures \cite{realspaceRG}, 
Migdal-Kadanoff block renormalizations \cite{MKRG} play a special role
because they can be considered in two ways, 
 either as approximate renormalization procedures on hypercubic lattices,
or as exact renormalization procedures on certain hierarchical lattices
\cite{berker,hierarchical}.
One of the most studied hierarchical lattice is the
diamond lattice which is constructed recursively
from a single link called here generation $n=0$ (see Figure \ref{figdiamond}):
 generation $n=1$ consists of $b$ branches, each branch
 containing $2$ bonds in series ;
 generation $n=2$ is obtained by applying the same transformation
to each bond of the generation $n=1$.
At generation $n$, the length $L_n$ between the two extreme sites
$A$ and $B$ is $L_n=2^n$, and the total number of bonds is 
\begin{eqnarray}
B_n=(2b)^n = L_n^{d_{Liens}(b)} \ \ \ {\rm \ \ with \  \ } 
d_{eff}(b)= \frac{ \ln (2b)}{\ln 2}
\label{dliens}
\end{eqnarray}
where $d_{eff}(b)$ represents some effective dimensionality.

On this diamond lattice, various
disordered models have been studied,
such as for instance the diluted Ising model \cite{diluted}, 
ferromagnetic random Potts model \cite{Kin_Dom,Der_Potts,andelman},
spin-glasses \cite{young,mckay,Gardnersg,bray_moore,nifle_hilhorst}
and the directed polymer model
 \cite{Der_Gri,Coo_Der,Tim,roux,kardar,cao, tang,Muk_Bha,Bou_Sil}.
In this article, we start from the exact renormalizations
existing for these disordered models on the diamond lattices
to derive the relations existing between the tails exponents 
and the droplet exponent $\omega$.

The paper is organized as follows.
The tails of the distribution $\Pi(x)$ of the ground state
energy of the directed polymer are discussed in 
Section \ref{sectiondp} together with numerical results;
the tails of the distribution ${\widetilde \Pi}(x)$ of excitations
in the ferromagnetic random Potts model and in the Ising spin-glasses
are studied in Section \ref{sectionpotts} and in Section \ref{sectionsg}
respectively.
Finally in Section \ref{sectiongene}, we generalize these results
to other lattices
via an analysis of the measure of the
rare disorder configurations governing the tails.
Our conclusions are summarized in Section \ref{conclusion}.
Appendix \ref{zerotemptails} contains more 
detailed calculations,
and Appendix \ref{zhangargument}
contains a brief reminder of Zhang argument for 
the directed polymer on hypercubic lattices.

\section{ Directed polymer on diamond lattice }

\label{sectiondp}

In this Section, we study the tails of the 
rescaled probability distribution $\Pi(x)$
for the ground state energy of the directed polymer model.
(Eq. \ref{scalinge0}).

\subsection{ Reminder on the exact renormalization } 

\subsubsection{ Renormalization at finite temperature}

The model of
a directed polymer in a random medium \cite{Hal_Zha}
can be defined on diamond hierarchical lattice with $b$ branches
 \cite{Der_Gri, Coo_Der,Tim,roux,kardar,cao, tang, Muk_Bha,Bou_Sil}.
The partition function
$Z_n$ of generation $n$ satisfies the exact recursion  \cite{Coo_Der}
\begin{eqnarray}
Z_{n+1} = \sum_{a=1}^b Z_n^{(2a-1)} Z_n^{(2a)}
\label{zrecursion}
\end{eqnarray}
where $(Z_n^{(1)},...,Z_n^{(2b)})$ are $(2b)$ independent partition functions
of generation $n$.
At generation $n=0$, the lattice reduces to a single bond
with a random energy $\epsilon$ drown from some distribution $\rho(\epsilon)$
and thus the initial condition for the recursion of Eq. \ref{zrecursion}
is simply
\begin{eqnarray}
Z_{n=0} = e^{- \beta \epsilon}
\label{zrecursioninitial}
\end{eqnarray}

In the low-temperature phase where the free-energy width $\Delta F(L)$
grows with the scale $L$, the recursion is dominated 
at large scale by the maximal term
 in Eq. \ref{zrecursion}
\begin{eqnarray}
Z_{n+1} \opsimeq max_{1 \leq a \leq b}
 \left( Z_n^{(2a-1)} Z_n^{(2a)} \right)
\label{recursionDPlowz}
\end{eqnarray}
or equivalently in terms of free-energies $F_n=-T \ln Z_n$
\begin{eqnarray}
F_{n+1} \opsimeq min_{1 \leq a \leq b}
 \left( F_n^{(2a-1)} + F_n^{(2a)} \right)
\label{recursionDPlow}
\end{eqnarray}
This effective low-temperature recursion 
coincides with the recursion of the energy $E_0$
of the ground state studied in \cite{Der_Gri,Coo_Der}.
The whole low-temperature phase is thus described by
the zero-temperature fixed-point.

\subsubsection{ Renormalization at zero temperature}

We now focus on the statistics of the ground state energy 
of the directed polymer \cite{Der_Gri,Tim,roux,tang}. 
At $T=0$, the recursion for the ground state energy 
involves the following minimisation \cite{Der_Gri}
\begin{eqnarray}
E_{n+1} = min\left[ E_n^{(1)}+ E_n^{(2)} ; E_n^{(3)}+ E_n^{(4)}; ...
; E_n^{(2b-1)}+ E_n^{(2b)}\right]
\label{egsrecursion}
\end{eqnarray}
This translates into the following recursion
for the probability $P_n(E)$ \cite{Der_Gri} :
\begin{eqnarray}
\int_z^{+\infty} dE P_{n+1}(E) = \left[ \int_z^{+\infty} dS Q_n(S) \right]^b
\label{egspdfrecursion}
\end{eqnarray}
where $Q_n(S)$ is the distribution of the sum $S=E_n^{(1)}+E_n^{(2)}$
\begin{eqnarray}
 Q_n(S) = \int_{-\infty}^{+\infty} dE_1 \int_{-\infty}^{+\infty} dE_2 
P_n(E_1) P_n(E_2) \delta( S- (E_1+E_2))
\label{Qnconvolution}
\end{eqnarray}

\subsubsection{ Renormalization in the scaling regime }

For large $n$, one expects the scaling \cite{Der_Gri}
\begin{eqnarray}
 P_n(E) \opsimeq_{n \to \infty} \frac{1}{ \delta_n} 
 \Pi_b \left( \frac{E-\gamma_n}{\delta_n} \right)
\label{pnescaling}
\end{eqnarray}
where the term $\gamma_n$ is extensive in the length $L_n$ 
\begin{eqnarray}
\gamma_n \opsimeq_{n \to \infty}  e_0 L_n 
\label{gamman}
\end{eqnarray}
i.e. $\gamma_{n+1}/\gamma_n \to 2$.
The width $\delta_n$ scales with the length $L_n$
with some a priori unknown exponent $\omega$ \cite{Der_Gri}
\begin{eqnarray}
\delta_n \opsimeq \lambda^n  \simeq L_n^{\omega} \ \ {\rm with \ \ } \lambda
 \equiv 2^{\omega}
\label{deltan}
\end{eqnarray}

Replacing the scaling form of Eq. \ref{pnescaling}
in the recursion of Eqs \ref{egspdfrecursion}, \ref{Qnconvolution}
yields
\begin{eqnarray}
\int_{\frac{z-\gamma_{n+1}}{\delta_{n+1}}}^{+\infty} dx \Pi_b(x) 
= \left[ \int_{\frac{z-2\gamma_{n}}{\delta_{n}}}^{+\infty} dx
 \int_{-\infty}^{+\infty} dx_1 \int_{-\infty}^{+\infty} dx_2 
\Pi_b(x_1)  \Pi_b(x_2) \delta( x- (x_1+x_2)) \right]^b
\label{recursionscal1}
\end{eqnarray}
Using $\gamma_{n+1}/\gamma_n=2$ and $\delta_{n+1}/\delta_n \to 
\lambda = 2^{\omega}$,
one obtains
\begin{eqnarray}
\int_{\frac{u}{\lambda}}^{+\infty} dx \Pi_b(x) 
= \left[ \int_{u}^{+\infty} dx G_b(x) \right]^b
\label{recursionscal}
\end{eqnarray}
where
\begin{eqnarray}
G_b(x) \equiv
 \int_{-\infty}^{+\infty} dx_1 \int_{-\infty}^{+\infty} dx_2 
\Pi_b(x_1)  \Pi_b(x_2) \delta( x- (x_1+x_2))
\label{gbconvolution}
\end{eqnarray}

The recursion simplifies in the limit $b=1$, where it reduces
to the Central Limit theorem for the sum of random variables
with  \cite{Der_Gri}
\begin{eqnarray}
 \omega(b=1) &&= \frac{1}{2} \nonumber \\
 \Pi_{b=1}(x) && = \frac{1 }{ \sqrt{2 \pi} } e^{- \frac{x^2}{2} }
\label{caseb1}
\end{eqnarray}
We refer to \cite{Der_Gri} where an expansion in $b=1+\epsilon$
has been developed.

Another limit where the recursion simplifies is $b \to \infty$.
In the limit, a single iteration consists in taking
the minimum of a large number $b \to \infty$ of random variables.
The rescaled distribution is then the Gumbel distribution \cite{Gum_Gal}
\begin{eqnarray}
 \omega(b \to \infty) &&= 0 \nonumber \\
 \Pi_{b \to \infty}(x) && = e^{x-e^x}
\label{casebinfty}
\end{eqnarray}

For $1<b<+\infty$, the exponent $\omega(b)$ is expected to decay from
$\omega(b=1)=1/2$ to $\omega(b=\infty)=0$, 
Accordingly, the rescaled distribution $\Pi_b(x)$ is expected
 to interpolate between the Gaussian and the Gumbel distribution.

\subsection{ Relations between tails exponents and the droplet exponent } 

\label{grounddptails}

Let us now focus on the tail exponents
 $\eta(b)$ and $\eta'(b)$ of the probability distribution $\Pi_b$
\begin{eqnarray}
\ln \Pi_b(x ) && \opsimeq_{x \to -\infty} - c (-x)^{\eta(b)} +...\nonumber \\
\ln \Pi_b(x ) && \opsimeq_{x \to +\infty} - d x^{\eta'(b)} +...
\label{defetamu}
\end{eqnarray}
From the two extreme cases of Eqs \ref{caseb1} and \ref{casebinfty},
one expects that the left exponent $\eta(b)$ varies between
\begin{eqnarray}
\eta(b=1) && =2  \nonumber \\
\eta(b=\infty) && =1
\label{limiteseta}
\end{eqnarray}
whereas the right exponent $\eta'(b)$  varies between
\begin{eqnarray}
\eta'(b=1) && =2  \nonumber \\
\eta'(b=\infty) && = \infty
\label{limitesmu}
\end{eqnarray}

The aim of this section is to show that these tail exponents 
are simply related to the droplet exponent $\omega(b)$ via
\begin{eqnarray}
 \eta(b) && = \frac{1}{1-\omega(b)} \nonumber \\
 \eta'(b) && = \frac{\ln (2b)}{\ln 2} \eta(b)
\label{resetamuomega}
\end{eqnarray}
To make things clearer, we have chosen to present here in the text
only simple saddle-point arguments
at leading order. We refer to Appendix  \ref{zerotemptails}
(see Eqs \ref{leadingsuite} and \ref{etamu})
for a much more detailed proof with subleading corrections.

\subsubsection{ Left-tail exponent $\eta$ }

Assume that the left-tail decay of the probability distribution $\Pi_b(x)$
is given at leading order by Eq. \ref{defetamu}.
A saddle-point analysis shows that the probability
distribution $G_b$ (Eq \ref{gbconvolution})
of the sum $x=x_1+x_2$
 presents the tail
\begin{eqnarray}
\ln G_b(x)  \opsimeq_{x \to -\infty}
 - 2 c \left( - \frac{x}{2} \right)^{\eta} +... 
\label{lefttail1}
\end{eqnarray}
The recursion of Eq. \ref{recursionscal} yields by differentiation that 
the tail of $\Pi_b$ is related to the tail of $G_b$ via
\begin{eqnarray}
\frac{1}{\lambda} \Pi_b \left( \frac{u}{\lambda}\right) 
\opsimeq_{u \to -\infty} b G_b(u)
 \left[ \int_{u}^{+\infty} dy G_b(y) \right]^{b-1} 
\opsimeq_{u \to -\infty} b G_b(u)  \oppropto_{u \to -\infty} 
b e^{ - 2 c \left( - \frac{u}{2} \right)^{\eta} +... }
\label{lefttail2}
\end{eqnarray}
Using $\lambda=2^{\omega}$, this yields in terms of the variable $x=u/\lambda$
\begin{eqnarray}
\ln  \Pi_b \left( x \right) 
\opsimeq_{x \to -\infty}  - 2 c \left( - \frac{\lambda x }{2} \right)^{\eta} +... 
=  - 2^{1+ \eta ( \omega-1)} c \left( -  x \right)^{\eta} +... 
\label{lefttailfin}
\end{eqnarray}
The consistency with the scaling form of the tail of Eq \ref{defetamu}
yields $\eta=1/(1-\omega)$ as
stated in Eq \ref{resetamuomega}.

\subsubsection{ Right-tail exponent $\eta'$ }

Assume that the right-tail decay of the probability distribution $\Pi_b(x)$
is given at leading order by (Eq. \ref{defetamu}).
A saddle-point analysis shows that the probability
distribution $G_b$ (Eq \ref{gbconvolution})
of the sum $x=x_1+x_2$
 presents the tail
\begin{eqnarray}
\ln G_b(x)  \opsimeq_{x \to +\infty}
 - 2 c \left(  \frac{x}{2} \right)^{\eta'} +... 
\label{righttail1}
\end{eqnarray}
The recursion of Eq. \ref{recursionscal} yields by differentiation that 
the tail of $\Pi_b$ is related to the tail of $G_b$ via
\begin{eqnarray}
\frac{1}{\lambda} \Pi_b \left( \frac{u}{\lambda}\right) 
\opsimeq_{u \to + \infty} b G_b(u)
 \left[ \int_{u}^{+\infty} dy G_b(y) \right]^{b-1} 
\oppropto_{u \to -\infty}  \left[ G_b(u) \right]^b \oppropto_{u \to -\infty} 
 e^{ - 2 b c \left( - \frac{u}{2} \right)^{\eta'} +... }
\label{righttail2}
\end{eqnarray}
Using $\lambda=2^{\omega}$, this yields in terms of the variable $x=u/\lambda$
\begin{eqnarray}
\ln  \Pi_b \left( x \right) 
\opsimeq_{x \to -\infty}  - 2  b c 
\left( - \frac{\lambda x }{2} \right)^{\eta'} +... 
=  - 2^{1+ \eta' ( \omega-1)} b  c \left( -  x \right)^{\eta'} +... 
\label{righttailfin}
\end{eqnarray}
The consistency with the scaling form of the tail of Eq \ref{defetamu}
yields $\eta'(1-\omega)= \ln (2b)/\ln 2$ as
stated in Eq \ref{resetamuomega}.

\subsubsection{ Discussion }

The relations of Eq. \ref{resetamuomega} have already been found for 
the special case $b=2$ in \cite{roux}.
However in \cite{roux}, a third relation between the three exponents
was also written, leading to the simple values
$\eta(b=2)=3/2$ and $\eta'(b=2)=3$
that seem now excluded numerically (see below).

\subsection{ Numerical results of the ground state energy statistics}

\label{numedp}

\subsubsection{ Method : numerical recursion of the probability distribution  }

\begin{figure}[htbp]
\includegraphics[height=6cm]{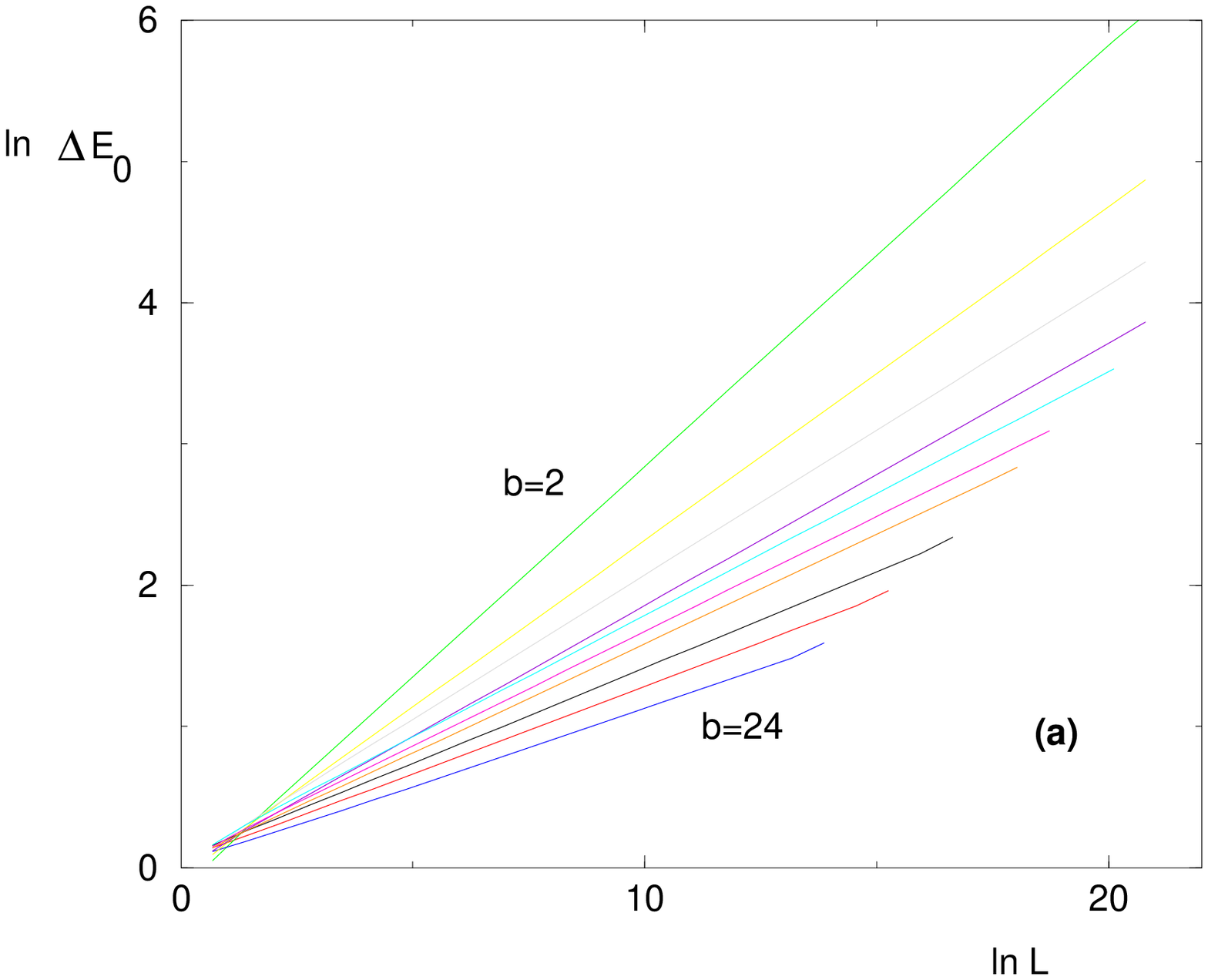}
\hspace{1cm}
\includegraphics[height=6cm]{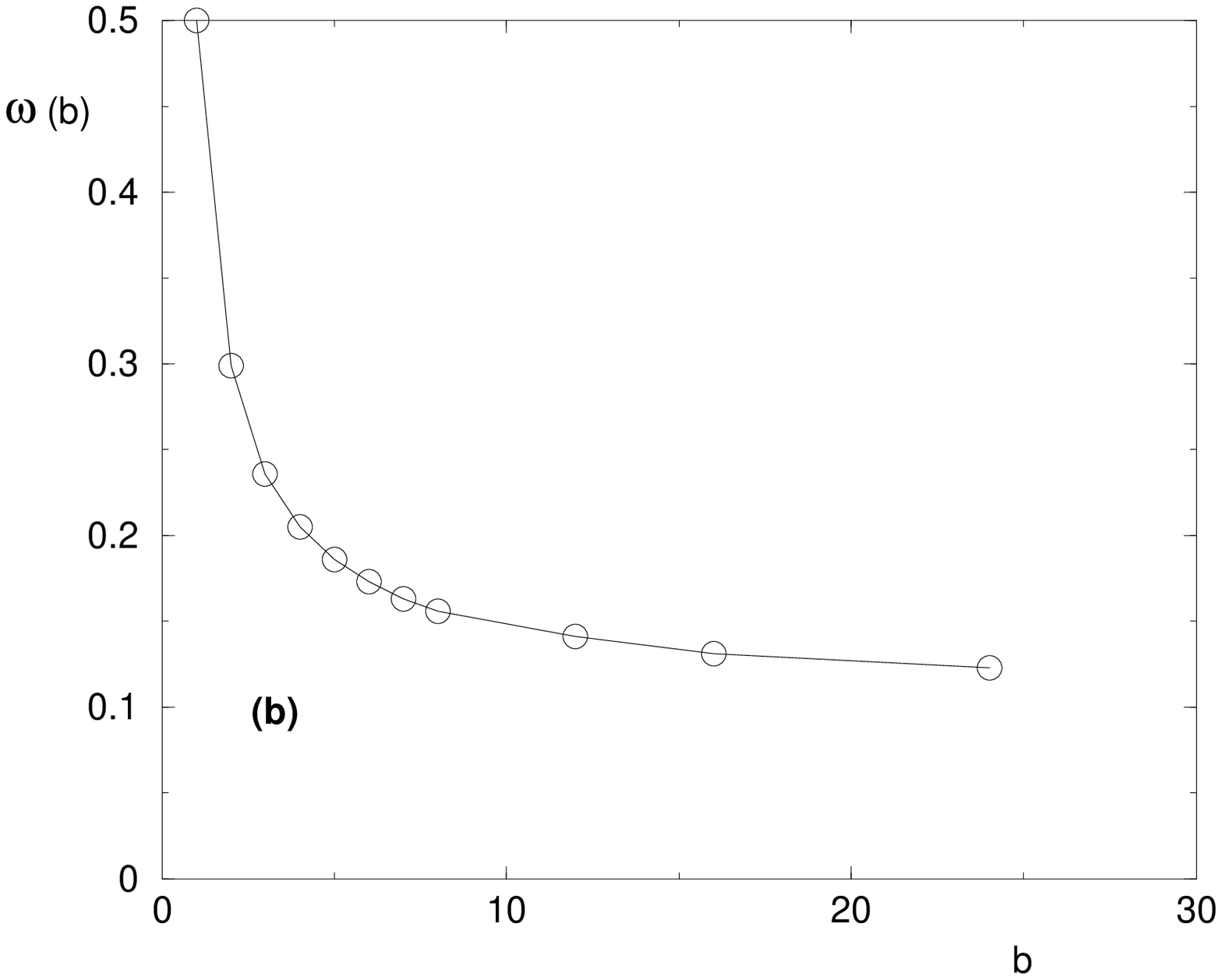}
\caption{(Color online) Width $\Delta E_0(L)$
of the ground state energy distribution for a polymer of length $L$
(a) the log-log plot of 
width $\Delta E_0 \sim L^{\omega(b)}$ for $b=2,3,4,5,6,7,8,12,16,24$.
(b) Exponent $\omega(b)$ as a function of the branching ratio $b$
(Precise numbers are given in Table \ref{table}) }
\label{figomegab}
\end{figure}

As explained in \cite{Der_Gri}, it is more convenient numerically
to consider the iteration of discrete probability
distributions of the form
\begin{eqnarray}
P_n(e) = \sum_{m=-\infty}^{+\infty} a_n(m) \delta_{e,m}
\label{pen}
\end{eqnarray}
where the energy $e$ can take only integer values.
This form is conserved via 
the recursion of Eq \ref{egsrecursion} that only involves
summation of energies and choice of minimal value.
The convolution step of Eq. \ref{Qnconvolution}
can be written as
\begin{eqnarray}
Q_n(e) \equiv (P_n*P_n)(e) = \sum_{m=-\infty}^{+\infty} b_n(m) \delta_{e,m}
\label{qen}
\end{eqnarray}
with the following rules
\begin{eqnarray}
 b_n(m) = \sum_{m'=-\infty}^{+\infty} a_n(m') a_n(m-m')
\label{bnm}
\end{eqnarray}
Since the function
\begin{eqnarray}
 \int_x^{+\infty} de Q_n(e) = \sum_{m > x } b_n(m)
=  \sum_{m=-\infty}^{+\infty} c_n(m) \theta( m -1 < x < m)  
\end{eqnarray}
is constant on intervals with values
\begin{eqnarray}
 c_n(m) = \sum_{m'=m}^{+\infty} b_n(m') 
\label{cnm}
\end{eqnarray}
it is easy to raise it to power $b$
\begin{eqnarray}
\left[ \int_x^{+\infty} de Q_n(e) \right]^b=
  \sum_{m'=-\infty}^{+\infty} \left[ c_n(m) \right]^b \theta( m -1 < x < m)  
\end{eqnarray}
Since we have
\begin{eqnarray}
 \int_x^{+\infty} de P_{n+1}(e) = \sum_{m > x } a_{n+1}(m)
=  \sum_{m'=-\infty}^{+\infty}
 \left[ \sum_{m'=m}^{+\infty} a_{n+1}(m')\right] \theta( m -1 < x < m)  
\end{eqnarray}
the recursion of Eq. \ref{egspdfrecursion} yields
\begin{eqnarray}
\sum_{m'=m}^{+\infty} a_{n+1}(m') = \left[ c_n(m) \right]^b
\end{eqnarray}
and thus the coefficients at generation $(n+1)$ can be obtained via
\begin{eqnarray}
 a_{n+1}(m) = \left[ c_n(m) \right]^b -  \left[ c_n(m+1) \right]^b
\label{anplus1zerot}
\end{eqnarray}

This method allows to obtain very accurate results for
the fluctuation exponent $\omega(b)$ and
for the tail exponents $\eta(b)$ and $\eta'(b)$
because the probability distribution can be evaluated
very far in the tails.
The results presented below have been obtained by the iteration
up to $n_g \sim 30$ generations
of the discrete distribution of Eq. \ref{pen}
with numerical tail cut-offs of order 
$- 10 000 \leq m - m_{center}(n) \leq 1000$
around the moving center $m_{center}(n)$ for each generation $n$.

\subsubsection{ Numerical results for the fluctuation exponent
$\omega(b)$  }

 \begin{table}[htbp]
 \centerline{
 \begin{tabular}{|l|l|l|}   \hline
 b  &     $ \ \ \omega(b) $ & $ \ \ \eta(b)$ \\ \hline
 2  &  \ 0.299  &   \ 1.43       \\ \hline
 3  &  \ 0.236  &   \ 1.31  \\ \hline
 4  &  \ 0.205  &   \ 1.26    \\ \hline
 5  &  \ 0.186  &   \ 1.23 \\ \hline
 6  &  \ 0.173  &   \ 1.21     \\ \hline
 7  &  \ 0.163 &    \ 1.20 \\ \hline
 8  &  \ 0.156  &   \ 1.18   \\ \hline
 12 &  \ 0.141 &    \ 1.16  \\ \hline
 16 &  \ 0.131  &   \ 1.15    \\ \hline
 24 &  \ 0.123 &    \ 1.14  \\ \hline
 \end{tabular}
 }

\caption{Results for the exponents $\omega(b)$ and $\eta(b)$ as 
the number $b$ of branches varies }
\label{table}
\end{table}

We first show on Fig. \ref{figomegab} the 
log-log plot of 
width $\Delta E_0 \sim L^{\omega(b)}$ of the ground state energy
probability distribution. The measures of the slopes yield
the values given in Table \ref{table}.
The values of $\omega(b)$ are in agreement with
the existing previous numerical measures \cite{Der_Gri,Tim,roux,tang}.
The curve $\omega(b)$ shown on Fig. \ref{figomegab} b
seems to suggest that the droplet exponent $\omega$ remains positive
as long as the effective dimension $d_{eff}(b)= \ln(2b)/\ln 2$ (Eq \ref{dliens})
remains finite.
Note that for hypercubic lattices, 
the existence of a finite upper critical dimension
$d_c$ above which the droplet exponent vanishes
has remained a very controversial issue between the numerical studies
\cite{Tan_For_Wol,Ala_etal,KimetAla,Mar_etal}
and various theoretical approaches \cite{Las_Kin,Col_Moo,LeDou_Wie}.

\begin{figure}[htbp]
\includegraphics[height=6cm]{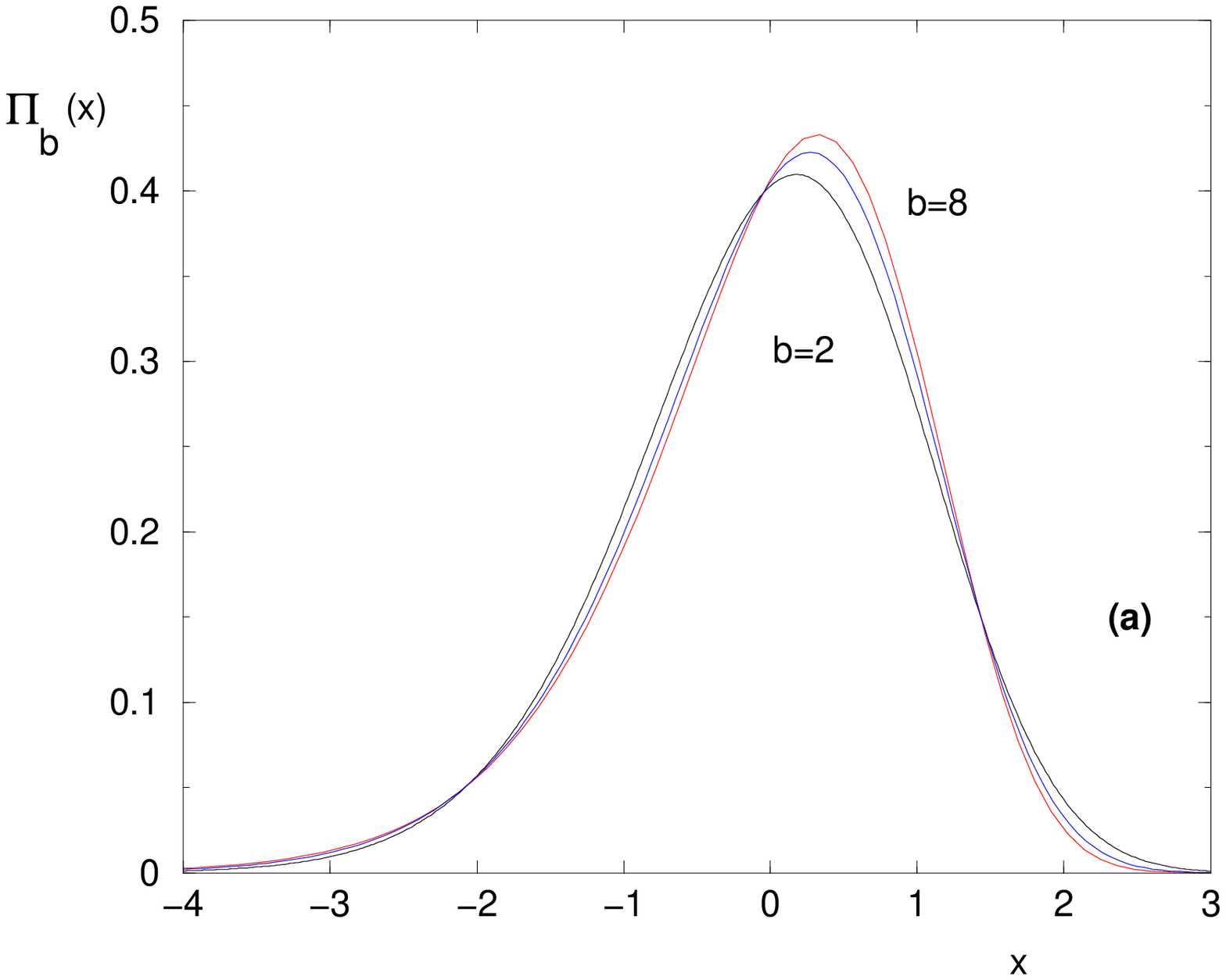}
\hspace{1cm}
\includegraphics[height=6cm]{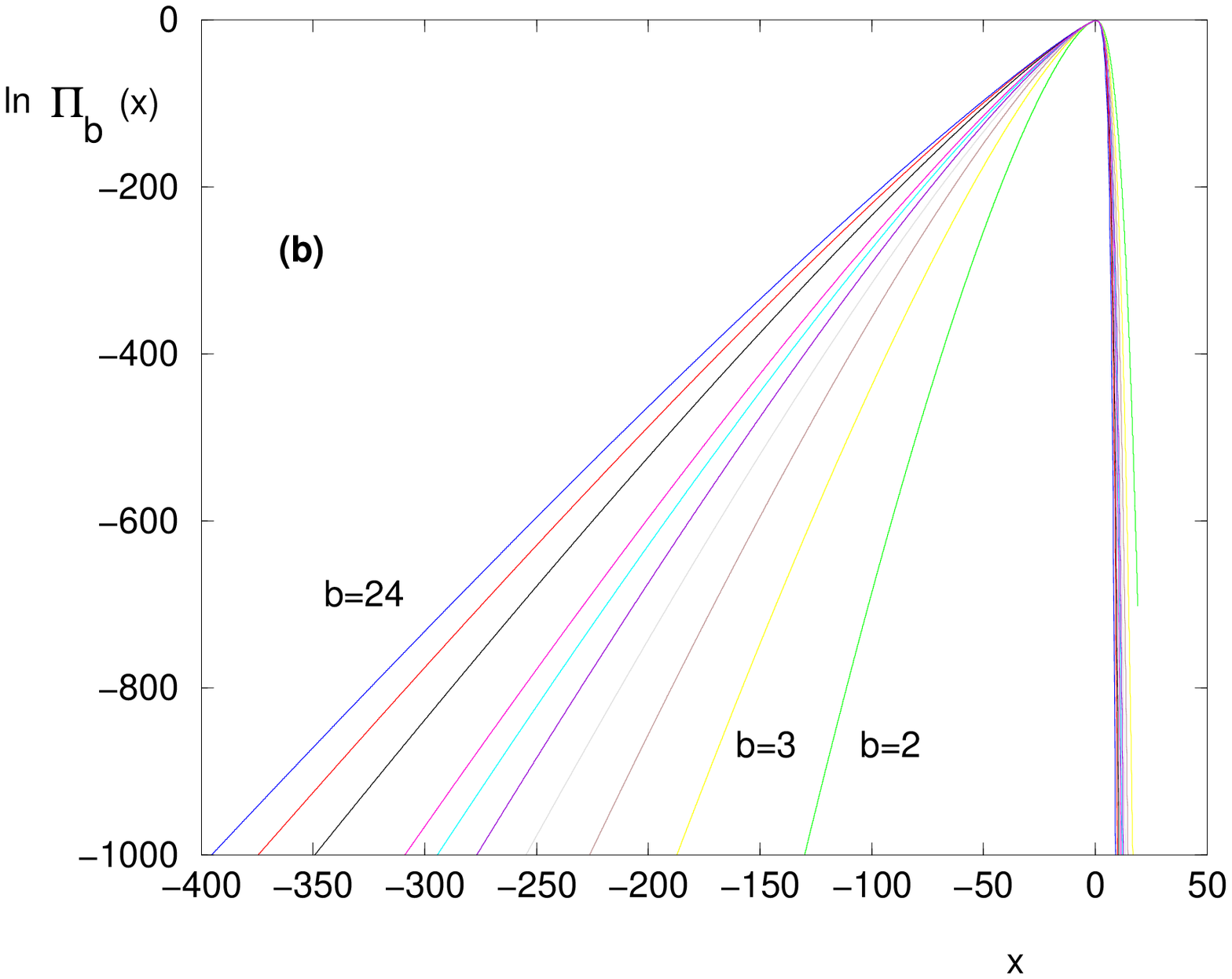}
\caption{
(Color online) Rescaled probability distribution of
the ground state energy 
(a) Bulk representation for $b=2,4,8$
(b) Log plot for $b=2,3,4,5,6,7,8,12,16,24$
 to see the behavior far in the tails
 }
\label{figpdfb}
\end{figure}

\begin{figure}[htbp]
\includegraphics[height=6cm]{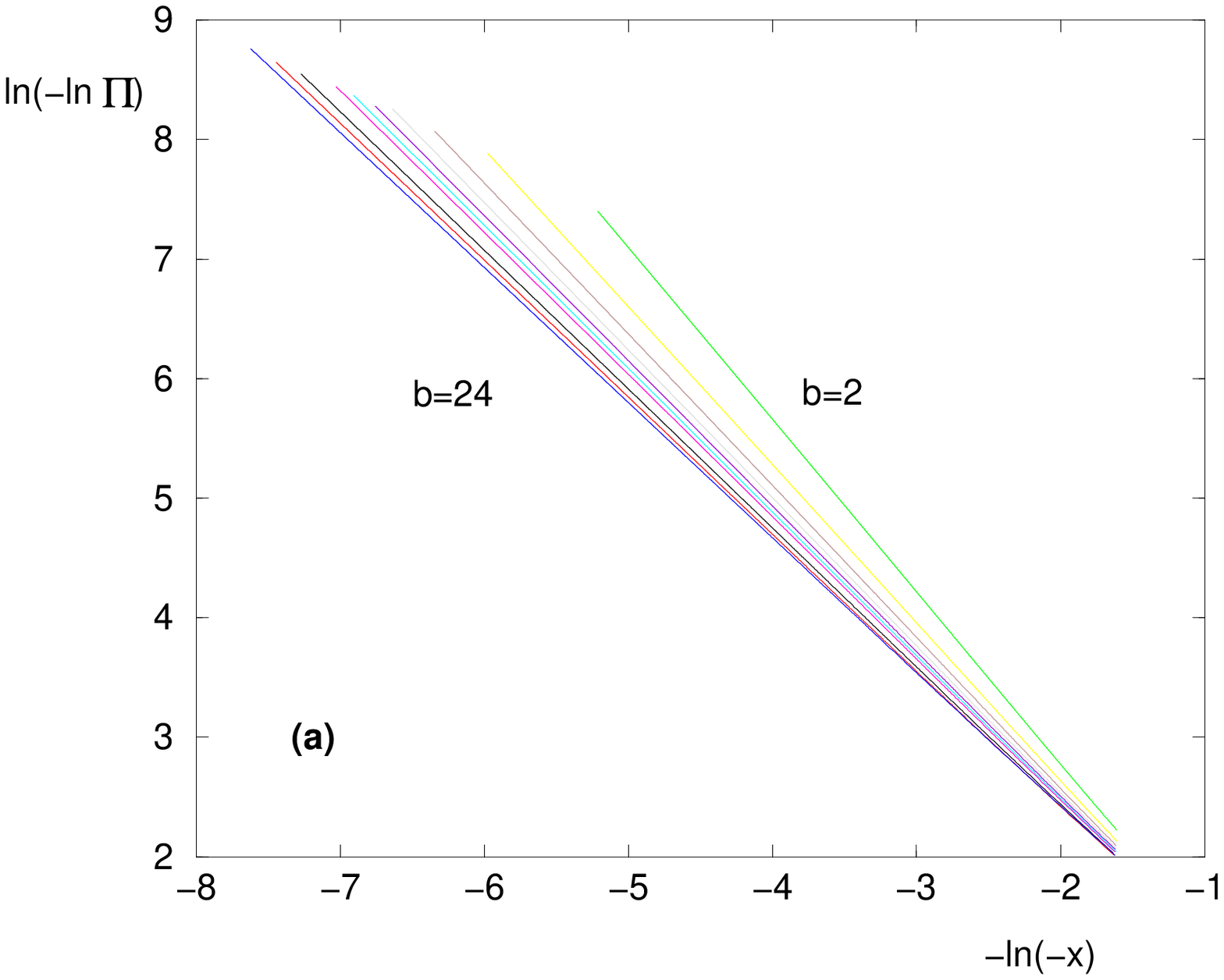}
\hspace{1cm}
\includegraphics[height=6cm]{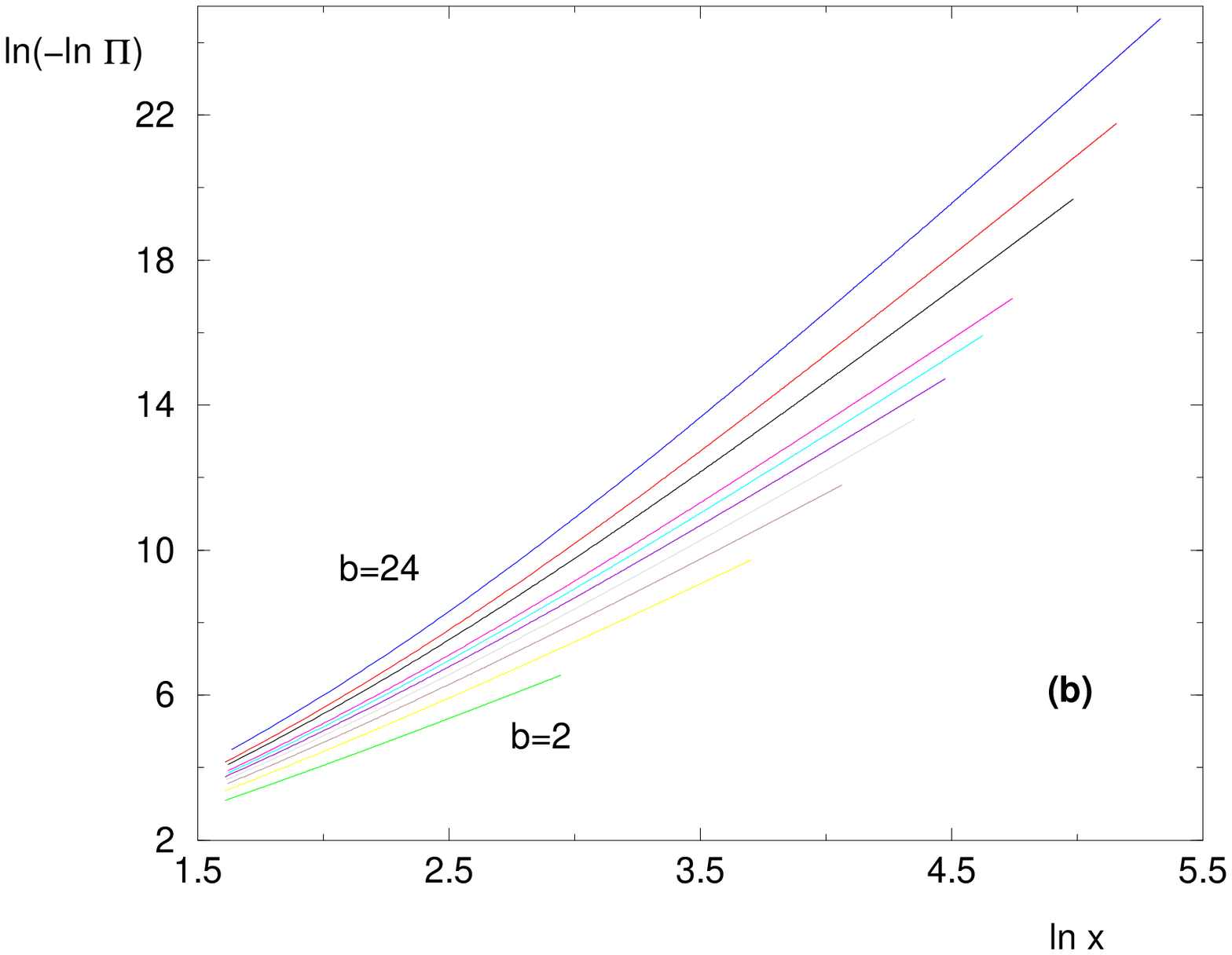}
\caption{
(Color online) Log-log plot of the tails
of the rescaled probability distribution of
the ground state energy for branching ratio $b=2,3,4,5,6,7,8,12,16,24$.
(a) Measure of the exponent $\eta(b)$ governing the left tail
$\ln \Pi_b(x \to -\infty)  \sim - (-x)^{\eta(b)}$
(b)  Measure of the exponent $\eta'(b)$ governing the right tail
$ \ln \Pi_b(x \to +\infty)  \sim - x^{\eta'(b)}$ 
The curvatures of $\ln (- \ln  \Pi_b(x))$ as a function of $\ln(x)$ 
show that the corrections to the leading behavior are stronger
than for the left tail shown in (a). }
\label{figtailsb}
\end{figure}

\subsubsection{ Numerical results for the rescaled probability distribution
$\Pi_b(x)$  }

We show on Fig. \ref{figpdfb} the asymptotic rescaled probability distributions
$\Pi_b(x)$ defined in Eq. \ref{scalinge0}.
On Fig. \ref{figpdfb} (a) we show $\Pi_b(x)$ in the bulk for $b=2,4,8$.
On  Fig. \ref{figpdfb} (b) we show $\ln \Pi_b(x)$ to 
 see the behaviors far in the tails.

From the Figures \ref{figtailsb}, we have measured the tails exponents
defined by Eq. \ref{defetamu}.
The left-tail exponent $\eta(b)$ can be measured as the slopes of
the curves of Fig. \ref{figtailsb} (a).
The results $\eta(b)$ are given in Table \ref{table}.
The relation $\eta(b)=1/(1-\omega(b))$ (Eq. \ref{resetamuomega})
 is well satisfied by
our numerical results.

The right-tail exponent $\eta'(b)$ turns out to be more difficult to measure precisely,
because the corrections to the leading behavior are much stronger,
as shown by the curvatures of our results of Fig.  \ref{figtailsb} (b).
However, the values predicted by the relation  of Eq. \ref{resetamuomega}
\begin{eqnarray}
\eta'(b)= \frac{ \ln (2b)}{\ln 2 } \ \eta(b)
\label{resmub}
\end{eqnarray}
are compatible with our data.

\subsection{ Extension to the critical point }

On the diamond hierarchical lattice, the free-energy fluctuations of
the directed polymer either grow as $L^{\omega}$ for $T<T_c$,
or decay as $L^{-\omega'}$ for $T>T_c$ or remain of order $O(1)$
exactly at $T_c$. The study of the tails of the critical rescaled distribution 
\cite{diamondpolymer} is a special case of the above relations 
(Eq \ref{resetamuomega})
with $\omega_c=0$\begin{eqnarray}
 \eta_c(b) && = 1 \nonumber \\
 \eta'_c(b) && = \frac{\ln (2b)}{\ln 2} 
\label{resetamucriti}
\end{eqnarray}
We refer to \cite{diamondpolymer} for more details.

\section{ Ising and Potts random ferromagnets on diamond lattice } 

\label{sectionpotts}

In this section, we study the tails of the rescaled distribution 
${\widetilde \Pi}(x)$ for the excitations in random Ising and Potts ferromagnets
(Eq. \ref{scalingexci}).

\subsection{ Reminder on the exact renormalization for the effective coupling }

The Ising Hamiltonian reads
\begin{eqnarray}
H_{Ising}= - \sum_{<i,j>}  J_{i,j} S_i S_j
\label{defising}
\end{eqnarray}
where the spins take the values $S_i=\pm 1$
and where the couplings $J_{i,j}$ are positive random variables
(see below section \ref{sectionsg} on spin-glass for couplings 
of random sign).
The Potts Hamiltonian is a generalization
where the variable $\sigma_i$ can take $q$ different values.
\begin{eqnarray}
H_{Potts}= - \sum_{<i,j>} 2 J_{i,j} \delta_{\sigma_i,\sigma_j}
\label{defpotts}
\end{eqnarray}
(We choose $(2J)$ to recover Ising for $q=2$)

\subsubsection{ Renormalization at finite temperature}

The effective coupling between the two end-points $A$ and $B$
of the diamond lattice of Fig. \ref{figdiamond}
is defined by
\begin{eqnarray}
e^{- 2 \beta J_{eff}} = \frac{Z_{+-}}{Z_{++}}
 = e^{- \beta \left( F_{+-} - F_{++}\right)}
\label{defjeff}
\end{eqnarray}
where $Z_{++}$ and $Z_{+-}$ are the partitions functions
corresponding respectively to the same color at both ends
or to two different colors at both ends. So the effective coupling
represents the free-energy cost of creating an interface between
the two ends at distance $L$.
The renormalization equation reads in terms of 
the variable $y=e^{2 \beta J}$ \cite{Kin_Dom,Der_Potts,andelman}
\begin{eqnarray}
y_{n+1}= \prod_{i=1}^b \left( \frac{ y_n^{(i_1)} y_n^{(i_2)} + (q-1) }
{y_n^{(i_1)} + y_n^{(i_2)} + (q-2)} \right)
\label{rgpotts}
\end{eqnarray}

In the low-temperature phase where the effective couplings $J_n$
grow with the length $L_n=2^n$, 
the contribution of each branch $(i)$
  is dominated by the minimal coupling between $(J_n^{(i_1)},J_n^{(i_2)})$.
 This leads to the effective 
zero-temperature recursion
\begin{eqnarray}
J_{n+1} \opsimeq \sum_{i=1}^b {\rm min} \left[J_n^{(i_1)},J_n^{(i_2)} \right]
\label{recursionpottslow}
\end{eqnarray}
The whole low-temperature phase is thus described by
the zero-temperature fixed-point.

\subsubsection{ Renormalization at zero temperature}

We now focus on the zero-temperature renormalization 
\begin{eqnarray}
J_{n+1} = \sum_{i=1}^b {\rm min} \left[J_n^{(i_1)},J_n^{(i_2)} \right]
\label{jzerotrecursion}
\end{eqnarray}
Note that two operations 'sum' and 'min' occur
in the opposite order with respect to the directed polymer case
of Eq. \ref{egsrecursion}.
Eq. \ref{jzerotrecursion} translates into the following recursion
for the probability $P_n(J)$  :
\begin{eqnarray}
P_{n+1}(J) =  \int_{0}^{+\infty} dK_1 P_n^{min}(K_1)
...   \int_{0}^{+\infty} dK_b P_n^{min}(K_b)\delta( J- (K_1+K_2+...+K_b))
\label{spinconvol}
\end{eqnarray}
where $P_n^{min}(K)$ is the distribution of the minimum
 $K=min(J_n^{(1)},J_n^{(2)})$ of two variables
 drawn with the distribution $P_n(J)$
\begin{eqnarray}
\int_z^{+\infty} dK P_n^{min}(K) = \left[ \int_z^{+\infty} dJ  P_n(J) \right]^2
\label{spinmin}
\end{eqnarray}

\subsubsection{ Renormalization in the scaling regime }

For large $n$, one expects the scaling 
\begin{eqnarray}
 P_n(J) \opsimeq_{n \to \infty} \frac{1}{ L_n^{\omega}} 
 {\widetilde \Pi}_b \left( \frac{ J  - \gamma_n }{L_n^{\omega}} \right)
\label{pnjscaling}
\end{eqnarray}

Replacing the scaling form of Eq. \ref{pnjscaling}
in the recursion of Eqs \ref{spinconvol}, \ref{spinmin}
yields
\begin{eqnarray}
\frac{1}{\lambda} {\widetilde \Pi}_b \left(\frac{u}{\lambda} \right)  = 
 \int_{-\infty}^{+\infty} dx_1 {\widetilde \Pi}_b^{min}(x_1)
...   \int_{-\infty}^{+\infty} dx_b {\widetilde \Pi}_b^{min}(x_b)\delta( u- (x_1+x_2+...+x_b))
\label{spinconvolscal}
\end{eqnarray}
and
\begin{eqnarray}
\int_z^{+\infty} dx {\widetilde \Pi}_b^{min}(x) = \left[ \int_z^{+\infty} dx  {\widetilde \Pi}_b(x) \right]^2
\label{spinminscal}
\end{eqnarray}
and where the term $\gamma_n$ should grow asymptotically as
 $\gamma_{n+1}/\gamma_n \to b$.
In terms of the effective dimension $d_{eff}(b)$ of Eq \ref{dliens},
this corresponds to 
\begin{eqnarray}
\gamma_n \sim b^n = L_n^{\frac{\ln b}{\ln 2}}
= L_n^{ d_{eff}(b)-1} 
\label{gammaninterface}
\end{eqnarray}
 as it should for an interface of dimension $d_s= d_{eff}(b)-1$.

The recursion simplifies in the limit $b \to \infty$ \cite{Gardnersg}
where a single iteration consists in summing a large number
of random variables
\begin{eqnarray}
 {\widetilde \Pi}_{b \to \infty }(x)= \frac{1}{\sqrt{2 \pi}} e^{- \frac{x^2}{2}}
\label{pnjbinfty}
\end{eqnarray}
(Note that here, the usual other simple limit $b=1$ is not interesting since 
it does not correspond to a positive droplet exponent).

\subsection{ Relation between the tail exponents and the droplet exponent }

Let us now focus on the tail exponents
 $\eta(b)$ and $\eta'(b)$ of the probability distribution ${\widetilde \Pi}_b$
defined as in Eq \ref{defetamu}.
The aim of this section is to show that these tail exponents 
are simply related to the droplet exponent $\omega(b)$ via
\begin{eqnarray}
 \eta(b) && = \frac{ \frac{\ln b }{\ln 2} }{\frac{\ln b}{\ln 2} - \omega(b) } \nonumber \\
 \eta'(b) && = \frac{ \frac{\ln (2b)}{\ln 2} }{\frac{\ln b}{\ln 2} - \omega(b) }
\label{resetamuomegaspins}
\end{eqnarray}
As in the directed polymer case, these relations
can be obtained via simple saddle-point arguments.

\subsubsection{ Left-tail exponent $\eta$ }

If the left tail of ${\widetilde \Pi}_b(x)$ is described by Eq \ref{defetamu},
the distribution ${\widetilde \Pi}_b^{min}$ 
presents the same decay by differentiation of Eq. \ref{spinminscal}
\begin{eqnarray}
 {\widetilde \Pi}_b^{min}(x) = 2 {\widetilde \Pi}_b(x)  \int_x^{+\infty} dy  {\widetilde \Pi}_b(y) 
\opsimeq_{x \to -\infty} 2 {\widetilde \Pi}_b(x)
\opsimeq_{x \to -\infty} 2 e^{- c (-x)^{\eta} }
\label{spinminscalleft}
\end{eqnarray}
A saddle-point analysis shows that 
the convolution of $b$ variables $K_i$
then presents the following decay (Eq \ref{spinconvolscal})
\begin{eqnarray}
\frac{1}{\lambda} {\widetilde \Pi}_b \left(\frac{u}{\lambda} \right)  
\opsimeq_{u \to -\infty} 
  e^{- b c \left(- \frac{u}{b} \right)^{\eta} }
\label{spinconvolscalleft}
\end{eqnarray}
Using $\lambda=2^{\omega}$, this yields in terms of the variable $x=u/\lambda$
\begin{eqnarray}
\ln {\widetilde \Pi}_b \left( x \right)  
\opsimeq_{x \to -\infty} 
  - b c \left(- \frac{\lambda x}{b} \right)^{\eta} 
\opsimeq_{x \to -\infty} 
 - b^{1-\eta} 2^{\eta \omega} c (-x)^{\eta}
\label{spinconvolscalleft2}
\end{eqnarray}
The consistency with the tail of Eq \ref{defetamu}
yields the following constraint $2^{\eta \omega} = b^{\eta-1}$
leading to the result for $\eta(b)$ given in Eq \ref{resetamuomegaspins}.

\subsubsection{ Right-tail exponent $\eta'$ }

If the right tail of ${\widetilde \Pi}_b(x)$ is described by Eq \ref{defetamu},
the distribution ${\widetilde \Pi}_b^{min}$ 
presents the following exponential decay 
by differentiation of Eq. \ref{spinminscal}
\begin{eqnarray}
 {\widetilde \Pi}_b^{min}(x) = 2 {\widetilde \Pi}_b(x)  \int_x^{+\infty} dy  {\widetilde \Pi}_b(y) 
\oppropto_{x \to + \infty}  {\widetilde \Pi}_b^2(x)
\oppropto_{x \to +\infty}  e^{- 2 d x^{\eta'} }
\label{spinminscalright}
\end{eqnarray}
A saddle-point analysis shows that 
the convolution of $b$ variables $K_i$
then presents the following decay (Eq \ref{spinconvolscal})
\begin{eqnarray}
\frac{1}{\lambda} {\widetilde \Pi}_b \left(\frac{u}{\lambda} \right)  
\opsimeq_{u \to +\infty} 
  e^{- 2 b d \left( \frac{u}{b} \right)^{\eta'} }
\label{spinconvolscalright}
\end{eqnarray}
Using $\lambda=2^{\omega}$, this yields in terms of the variable $x=u/\lambda$
\begin{eqnarray}
\ln {\widetilde \Pi}_b \left( x \right)  
\opsimeq_{x \to +\infty} 
  - 2 b d \left( \frac{\lambda x}{b} \right)^{\eta'} 
\opsimeq_{x \to + \infty} 
 - b^{1-\eta'} 2^{1+\eta' \omega} d x^{\eta'}
\label{spinconvolscalright2}
\end{eqnarray}
The consistency with the tail of Eq \ref{defetamu}
yields the following constraint $2^{1+\eta' \omega} = b^{\eta'-1}$
leading to the result for $\eta'(b)$ given in Eq \ref{resetamuomegaspins}.

\subsection{ Extension to the critical point }

As in the directed polymer case (Eq \ref{resetamucriti}),
one may try to extend the results on the tail behaviors in the low-temperature
phase where $\omega>0$ to the critical point where $\omega_c=0$.
The right tail exponent becomes at criticality (Eq \ref{resetamuomegaspins})
\begin{eqnarray}
 \eta_c'(b) && = \frac{ \ln (2b)}{\ln b}
\label{resmucritispins}
\end{eqnarray}
For the left tail however, the problem is qualitatively different at
criticality and will not be discussed here
( the critical invariant distribution $P_c(J)$
does not extend to $(-\infty)$ anymore, 
but reaches the natural boundary $J=0$.)

\section{ Spin-glasses on diamond lattice }

\label{sectionsg}

We now consider the case of Ising spin-glasses described by Eq. \ref{defising},
but now the random couplings  $J_{i,j}$ can be positive or negative
and are distributed with a symmetric 
distribution $P_0(J)$ 
\begin{eqnarray}
P_0(J)=P_0(-J)
\label{defsg}
\end{eqnarray}

\subsection{ Reminder on the exact renormalization for the effective coupling }

\subsubsection{ Renormalization at zero temperature}

The renormalization at finite temperature
is still described by Eq. \ref{rgpotts} with $q=2$.
However, the fact that the couplings are of arbitrary sign
yields that 
the zero-temperature renormalization now reads
\begin{eqnarray}
J_{n+1} = \sum_{i=1}^b  {\rm sign} \left(J_n^{(i_1)}J_n^{(i_2)} \right)
{\rm min} \left[\vert J_n^{(i_1)}\vert ,\vert J_n^{(i_2)} \vert \right]
\label{sgzerotrecursion}
\end{eqnarray}
instead of Eq \ref{jzerotrecursion} corresponding to the case
 of positive couplings only.
The symmetry of the initial condition (Eq \ref{sgzerotrecursion})
is conserved by the renormalization
\begin{eqnarray}
P_n(J)=P_n(-J)
\label{sgsympn}
\end{eqnarray}
To translate the renormalization of Eq. \ref{sgzerotrecursion}
into a recursion for the probability distribution $P_n(J)$,
it is convenient to introduce the 
probability distribution $P_n^{abs}$ of $\vert J_n \vert$
\begin{eqnarray}
P_n^{abs}(J>0) =P_n(J)+P_n(-J)= 2 P_n(J)
\label{pnabs}
\end{eqnarray}
and the probability distribution
$Q_n^{abs}(K)$ of 
$K={\rm min} \left[\vert J_n^{(1)}\vert ,\vert J_n^{(2)} \vert \right] \geq 0$
satisfying
\begin{eqnarray}
\int_z^{+\infty} dK Q_n^{abs}(K) 
= \left[ \int_z^{+\infty} dJ  P_n^{abs}(J) \right]^2
\label{sgmin}
\end{eqnarray}
(the value $z=0$ corresponds to the normalization condition of both distribution).
The probability distribution $Q_n(J)$ of 
$K={\rm sign} \left(J_n^{(1)}J_n^{(2)} \right)
{\rm min} \left[\vert J_n^{(1)}\vert ,\vert J_n^{(2)} \vert \right] \geq 0$
is symmetric in $K \to -K$
\begin{eqnarray}
Q_n(K \neq 0) = \frac{1}{2} Q_n^{abs} \left( \vert K \vert \right)
\label{sgqnqnabs}
\end{eqnarray}
The recursion of Eq. \ref{sgzerotrecursion}
corresponds to the convolution
\begin{eqnarray}
P_{n+1}(J) =  \int_{-\infty}^{+\infty} dK_1 Q_n(K_1)
...  \int_{-\infty}^{+\infty} dK_b Q_n(K_b)  \delta( J- (K_1+K_2+...+K_b))
\label{sgconvol}
\end{eqnarray}

\subsubsection{ Renormalization in the scaling regime }

For large $n$, one expects the scaling 
\begin{eqnarray}
 P_n(J) \opsimeq_{n \to \infty} \frac{1}{ L_n^{\omega}} 
 {\widetilde \Pi}_b \left( \frac{ J }{L_n^{\omega}} \right)
\label{pnsgscaling}
\end{eqnarray}
where $\gamma_n=0$ in contrast with the ferromagnetic case of Eq \ref{pnjscaling},
and where the scaling function 
${\widetilde \Pi}_b$ is symmetric ( Eq. \ref{sgsympn} )
\begin{eqnarray}
{\widetilde \Pi}_b(x) = {\widetilde \Pi}_b(- x)
\label{pibsymm}
\end{eqnarray}

\subsection{ Relation between the tail exponent and the droplet exponent }

Note that the {\it true} rescaled distribution 
${\widetilde \Pi}^{true}(x)$ as defined by Eq. \ref{scalingexci}
with a positive energy for excitations above the ground state
actually corresponds to the distribution of the
absolute value of the rescaled coupling ${\widetilde \Pi}_b( x)$
defined in Eq \ref{pnsgscaling}
\begin{eqnarray}
{\widetilde \Pi}^{true}_b(x) = {\widetilde \Pi}_b( x)+ {\widetilde \Pi}_b( -x)
\label{pibtrue}
\end{eqnarray}
So the true distribution ${\widetilde \Pi}^{true}_b(x)$ begins at $x=0$
with a non-zero value ${\widetilde \Pi}^{true}_b(0)$ \cite{Fis_Hus}
and there exists a single tail as $x \to +\infty$.
So in contrast with the previous cases of the directed polymer
and of the ferromagnetic Potts model described
 by two tail exponents (Eq \ref{defetamu}),
the distribution ${\widetilde \Pi}_b(x)$ 
of the rescaled coupling for spin-glasses 
presents a single exponent 
 $\eta'(b)$ as a consequence of the symmetry of Eq. \ref{pibsymm}
\begin{eqnarray}
\ln {\widetilde \Pi}_b(x) \opsimeq_{ x \to \pm \infty} -d \vert x \vert^{\eta'}
\label{defetasg}
\end{eqnarray}
We will not repeat here the calculations that are similar to the
case of the right tail of the Potts model and that lead to the same
relation between exponents (Eq \ref{resetamuomegaspins})
\begin{eqnarray}
 \eta'(b)  = \frac{ \frac{\ln (2b)}{\ln 2} }{\frac{\ln b}{\ln 2} - \omega(b) }
\label{resetamuomegaspinssg}
\end{eqnarray}

\section{ Generalization of the tail exponents relations to other lattices }

\label{sectiongene}

In the previous sections, we have derived the relations that exist
between the tail exponents $(\eta,\eta')$ and the scaling exponent $\omega$
for various disordered models on the diamond hierarchical lattices
from the exact renormalization recursions.
The obtained relations are actually very simple in terms of the effective dimension
$d_{eff}(b)$ of these lattices (Eq. \ref{dliens}).
This suggest that these relations should have a simple interpretation.
In this section, we explain the physical meaning of these relations
and generalize them to other lattices.

\subsection{ Tail exponents for the directed polymer }

\subsubsection{ Physical interpretation of the left tail}

The relation $\eta=1/(1-\omega)$ ( Eq. \ref{resetamuomega} )
derived previously from the exact renormalization
on the diamond lattice can be interpreted as follows.
The left tail of the ground state energy of the directed polymer
corresponds to samples that leads to much lower energy than the average.
Let us evaluate the probability to obtain a ground state energy 
$E_0=\gamma_n - a L_n$ extensively below the averaged value $\gamma_n$
of the scaling function of Eq. \ref{pnescaling} with tail behavior
described by Eq \ref{defetamu}
\begin{eqnarray}
P_n(E_0=\gamma_n-a L_n) \simeq  \frac{1}{ L_n^{\omega} } 
 \Pi_b \left( - a \frac{L_n}{L_n^{\omega}} \right)
\oppropto_{L_n \to \infty}
e^{- c a^{\eta} L_n^{\eta (1-\omega)} }
\label{rarescaldpleft}
\end{eqnarray}
On the other hand, to obtain such a ground state energy 
$E_0=\gamma_n - a L_n$ extensively below the averaged value $\gamma_n$,
it seems reasonable to ask that each bond of the ground state
configuration of length $L$ should have an energy $\epsilon_i$
lower than the average, which happens with the exponentially small probability
\begin{eqnarray}
\left[ \int_{-\infty}^{..} d\epsilon \rho(\epsilon) \right]^L \sim e^{- cst L}
\label{raredpleft}
\end{eqnarray}
The identification of the length exponents in Eqs \ref{rarescaldpleft}
and \ref{raredpleft} yields
\begin{eqnarray}
\eta(1-\omega)=1
\label{etaleftraredp}
\end{eqnarray}
which corresponds to the relation found for the diamond lattice 
( Eq. \ref{resetamuomega}).
From this interpretation in terms of the rare disordered samples
that govern the left tail of the ground state energy configuration,
we expect that the relation of Eq \ref{etaleftraredp} is actually
valid on any lattice, and in particular for hypercubic lattices
of $1+d$ dimensions.
The relation is satisfied by the exact exponents in $1+1$ dimensions
with $\omega=1/3$ 
\cite{Hus_Hen_Fis,Kar,Joh,Pra_Spo}  and $\eta=3/2$ \cite{Joh,Pra_Spo,prae}.
The relation of Eq \ref{etaleftraredp}
has been previously derived for hypercubic lattices
via Zhang argument \cite{Hal_Zha} 
(the argument is recalled in Appendix 
\ref{zhangargument} for comparison)
and has been checked numerically in \cite{DPlefttail} for $d=2,3$.
However Zhang argument only concerns the left tail 
because it is based on the existence of a Lyapunov exponent
for positive moments of the partition function.
It cannot be extended easily to the right tail
that would be in correspondence with negative moments.
This is in contrast with the rare events analysis presented here 
that can be extended
 to the right tail as we now explain.

\subsubsection{ Physical interpretation of the right tail}

The relation $\eta'=d_{eff}(b)/(1-\omega)$ ( Eq. \ref{resetamuomega} )
derived previously from the exact renormalization
on the diamond lattice can be interpreted as follows.
The right tail of the ground state energy of the directed polymer
corresponds to samples that leads to much higher energy than the average.
Let us evaluate the probability to obtain a ground state energy 
$E_0=\gamma_n + a L_n$ extensively higher the averaged value $\gamma_n$
of the scaling function of Eq. \ref{pnescaling} with tail behavior
described by Eq \ref{defetamu}
\begin{eqnarray}
P_n(E_0=\gamma_n+a L_n) \simeq  \frac{1}{ L_n^{\omega} } 
 \Pi_b \left(  a \frac{L_n}{L_n^{\omega}} \right)
\oppropto_{L_n \to \infty}
e^{- d a^{\eta'} L_n^{\eta' (1-\omega)} }
\label{rarescaldpright}
\end{eqnarray}
On the other hand, to obtain such a ground state energy 
$E_0=\gamma_n + a L_n$ extensively higher the averaged value $\gamma_n$,
it seems reasonable to ask that all $L^{d_{eff}}$
bonds of the lattice should have an energy $\epsilon_i$
higher than the average, which happens with the exponentially small probability
\begin{eqnarray}
\left[ \int_{...}^{+\infty} d\epsilon \rho(\epsilon) \right]^{L^{d_{eff}}} \sim 
e^{- cst L^{d_{eff}}}
\label{raredpright}
\end{eqnarray}
The identification of the length exponents in Eqs \ref{rarescaldpright}
and \ref{raredpright}  
\begin{eqnarray}
\eta'(1-\omega)=d_{eff}
\label{etarightraredp}
\end{eqnarray}
exactly corresponds to the relation found for the diamond lattice 
( Eq. \ref{resetamuomega}).
From this interpretation in terms of the rare disordered samples,
we expect that the relation of Eq \ref{etarightraredp} is actually
valid on any lattice, and in particular for hypercubic lattices
of $d_{eff}=1+d$ dimensions.
The relation is satisfied by the exact exponents in $1+1$ dimensions
with $\omega=1/3$ 
\cite{Hus_Hen_Fis,Kar,Joh,Pra_Spo}  and $\eta'=3$ \cite{Joh,Pra_Spo,prae}.

\subsection{ Tail exponents for the ferromagnetic random Potts model }

\subsubsection{ Physical interpretation of the left tail}

The relation derived for the left tail of the Potts model on the diamond
lattice (Eq \ref{resetamuomegaspins}) reads in terms
of the effective dimension $d_{eff}(b)$ of Eq \ref{dliens}
\begin{eqnarray}
 \eta(b)  = \frac{d_{eff}(b)- 1  }{d_{eff}(b)-1 - \omega(b) } 
\label{leftpottsdeff}
\end{eqnarray}
We now propose the following physical interpretation.
The left tail 
corresponds to samples that leads to much lower effective
coupling than the average $\gamma_n \sim L_n^{d_{eff}(b)-1} $ 
(Eq. \ref{gammaninterface})
Let us evaluate the probability to obtain an effective coupling 
$J=\gamma_n - a  L_n^{d_{eff}(b)-1} $ 
extensively below the averaged value $\gamma_n$
of the scaling function of Eq. \ref{pnjscaling} with tail behavior
described by Eq \ref{defetamu}
\begin{eqnarray}
P_n(J=\gamma_n-a L_n^{d_{eff}(b)-1}) \simeq  \frac{1}{ L_n^{\omega} } 
 {\widetilde \Pi}_b \left( - a \frac{L_n^{d_{eff}(b)-1}}{L_n^{\omega}} \right)
\oppropto_{L_n \to \infty}
e^{- c a^{\eta} L_n^{\eta (d_{eff}(b)-1-\omega)} }
\label{rarescalpottsleft}
\end{eqnarray}
On the other hand, to obtain such a low effective coupling 
$J=\gamma_n - a  L_n^{d_{eff}(b)-1}$ extensively below the 
averaged value $\gamma_n$,
it seems reasonable to ask that each bond of the interface
of dimension $L^{d_{eff}(b)-1}$ should have a coupling $J_i$
lower than the average, which happens with the exponentially small probability
\begin{eqnarray}
\left[ \int_{0}^{..} dJ P_0(J) \right]^{L^{d_{eff}(b)-1}}
 \sim e^{- cst L^{d_{eff}(b)-1}}
\label{rarepottsleft}
\end{eqnarray}
The identification of the length exponents in Eqs \ref{rarescalpottsleft}
and \ref{rarepottsleft}  
\begin{eqnarray}
\eta ({d_{eff}(b)-1-\omega)} = d_{eff}(b)-1
\label{etaleftrarepotts}
\end{eqnarray}
exactly corresponds to the relation found for the diamond lattice 
( Eq. \ref{leftpottsdeff}).

\subsubsection{ Physical interpretation of the right tail}

The relation derived for the left tail of the Potts model on the diamond
lattice (Eq \ref{resetamuomegaspins}) reads in terms
of the effective dimension $d_{eff}(b)$ of Eq \ref{dliens}
\begin{eqnarray}
 \eta'(b)  = \frac{ d_{eff}(b) }{ d_{eff}(b)-1- \omega(b) }
\label{rightpottsdeff}
\end{eqnarray}
We now propose the following physical interpretation.
The right tail 
corresponds to samples that leads to much higher effective
coupling than the average $\gamma_n \sim L_n^{d_{eff}(b)-1} $ 
(Eq. \ref{gammaninterface})
Let us evaluate the probability to obtain an effective coupling 
$J=\gamma_n + a  L_n^{d_{eff}(b)-1} $ 
extensively above the averaged value $\gamma_n$
of the scaling function of Eq. \ref{pnjscaling} with tail behavior
described by Eq \ref{defetamu}
\begin{eqnarray}
P_n(J=\gamma_n+a L_n^{d_{eff}(b)-1}) \simeq  \frac{1}{ L_n^{\omega} } 
 {\widetilde \Pi}_b \left(  a \frac{L_n^{d_{eff}(b)-1}}{L_n^{\omega}} \right)
\oppropto_{L_n \to \infty}
e^{- d a^{\eta'} L_n^{\eta' (d_{eff}(b)-1-\omega)} }
\label{rarescalpottsright}
\end{eqnarray}
On the other hand, to obtain such a high effective coupling 
$J=\gamma_n + a  L_n^{d_{eff}(b)-1}$ extensively above the 
averaged value $\gamma_n$,
it seems reasonable to ask that all bonds of the sample
of dimension $L^{d_{eff}(b)}$ should have a coupling $J_i$
higher than the average, which happens with the exponentially small probability
\begin{eqnarray}
\left[ \int_{..}^{+\infty} dJ P_0(J) \right]^{L^{d_{eff}(b)}}
 \sim e^{- cst L^{d_{eff}(b)}}
\label{rarepottsright}
\end{eqnarray}
The identification of the length exponents in Eqs \ref{rarescalpottsright}
and \ref{rarepottsright}  
\begin{eqnarray}
\eta' ({d_{eff}(b)-1-\omega)} = d_{eff}(b)
\label{etarightrarepotts}
\end{eqnarray}
exactly corresponds to the relation found for the diamond lattice 
( Eq. \ref{rightpottsdeff}).

\section{ Conclusion} 

\label{conclusion}

In this paper, we have studied the statistics of excitations
 of finite-dimensional random models
(directed polymer, ferromagnetic random Potts model, Ising spin-glasses)
in their low-temperature phase
characterized by a positive droplet exponent $\omega>0$.
We have shown that the tails of the rescaled probability distribution
are characterized by two tails exponents $(\eta,\eta')$
that are simply related to the droplet exponent $\omega$.
We have first proved these relations 
 on the diamond hierarchical 
lattices where exact renormalizations exist for the rescaled probability 
distribution. We have then given the physical meaning
of these relations in terms of the measure of the
rare disorder configurations governing the tails.
This interpretation allows to understand the asymmetry $\eta<\eta'$
because a 'good' sample contributing to the left tail
is a sample containing $L^{d_s}$ 'good' random variables
for an interface of dimension $d_s<d$,
whereas a 'bad' sample contributing to the right tail
is a sample containing $L^{d}$ 'bad' random variables
for the bulk of dimension $d$.
We have then argued that this physical interpretation
means that the relations between the tails exponent $(\eta,\eta')$
and the droplet exponent $\omega$ should actually remain true
on arbitrary lattices when expressed in terms of the dimensions $(d_s,d)$,
namely 
\begin{eqnarray}
\eta  && =  \frac{d_s}{d_s-\omega} \nonumber \\
\eta' && = \frac{d}{d_s-\omega}
\label{eqconclusion}
\end{eqnarray}
The directed polymer corresponds to the case of
a linear object $d_s=1$ embedded in a space of total dimension $d$,
whereas the ferromagnetic random Potts model
corresponds to an interface of dimension $d_s=d-1$ in a space of dimension $d$.
These two cases merge for the special case of a linear object $d_s=1$ 
embedded in a space of total dimension $d=2$, which is not surprising
since the directed polymer model was precisely invented to model
a one-dimensional interface in two-dimensional ferromagnetic spin models
at low temperature \cite{Hus_Hen}. This special case also explains
why it is the distribution of
the ground state energy of the directed polymer
model (Eq. \ref{scalinge0}) which is in direct correspondence
with the distribution of excitations in ferromagnetic spin models
(Eq. \ref{scalingexci}).

 The case of spin-glasses is different for at least two reasons.
First of all, only the right tail exponent $\eta'$ exists, because the
distribution 
of the energy of excitations extends down to $E=0$
as a consequence of the symmetry $J \to -J$
(see the discussion around Eqs \ref{pibtrue} and \ref{defetasg}).
Secondly, in real space, the dimension $d_s$ 
is expected to be different from the value $(d-1)$
and to reflect the fractal nature of the droplet boundary 
\cite{Fis_Hus}.

As a final remark, we should stress that the tails exponents discussed
here concern the universal scaling distributions 
of the rescaled variables.
But of course, as in the Central Limit theorem, 
non-universal tails could also be present in random systems
with particular initial disorder distributions.

\appendix

\section{Tail exponents of the ground state energy distribution
for the directed polymer on the diamond lattice}

\label{zerotemptails}

In this Appendix, we derive the relations between
these tail exponents $\eta(b),\eta'(b)$ defined in Eqs \ref{defetamu}
and the fluctuation exponent $\omega(b)$.
We start from the recursion Eqs \ref{recursionscal1} and \ref{recursionscal}
in the scaling regime.
The convolution relation of Eq. \ref{gbconvolution}
is simple in Fourier
\begin{eqnarray}
 {\hat G}_b(q) = \left[{\hat F}_b(q) \right]^2
\label{convolfourier}
\end{eqnarray}
with
\begin{eqnarray}
 {\hat G}_b(q) &&\equiv \int_{-\infty}^{+\infty} dx e^{i k x} G_b(x) 
\ \ \ \ \ \ \  
G_b(x) = \int_{-\infty}^{+\infty} \frac{dq}{2  \pi} e^{-i q x} {\hat G}_b(q)
\nonumber \\
 {\hat F}_b(q) && \equiv \int_{-\infty}^{+\infty} dx e^{i q x} \Pi_b(x) 
\ \ \ \ \ \ \  
\Pi_b(x) = \int_{-\infty}^{+\infty} \frac{dq}{2  \pi} e^{-i q x} {\hat F}_b(q)
\label{fourier}
\end{eqnarray}
but the recursion relation Eq. \ref{recursionscal} 
is non-local in Fourier. This is why it is difficult to obtain
an explicit solution for the probability distribution $\Pi_b$.
In the following, we show that the problem simplifies for the tails
of the distribution $\Pi_b$.

\subsection{ Study of the left-tail form $\Pi_b(x \to -\infty)$ }

We write the left-tail of $\Pi_b(x)$ as
\begin{eqnarray}
 \Pi_b(x) \opsimeq_{x \to - \infty}
  e^{- c  \vert x \vert^{\eta} + \Phi
\left( \vert x \vert \right) }
\label{fblefttail}
\end{eqnarray}
where the function $\Phi(\vert x \vert)$ is subleading 
with respect to the first term of order $\vert x \vert^{\eta}$.
This left-tail will determine the asymptotic of the Fourier transform
${\hat F}_b(q)$ of Eq. \ref{fourier} for $q=i s$ with real $s \to +\infty$
\begin{eqnarray}
{\hat \Pi_b}(q=is) = \int_{-\infty}^{+\infty} dx e^{ - s x } \Pi_b(x)
\opsimeq_{s \to + \infty} \int dx e^{  s x 
- c x^{\eta}+ \Phi(x)  } 
\label{pns}
\end{eqnarray}
Since $\Phi(x)$ is subleading, 
we perform a saddle point calculation with the two first terms,
yielding the saddle value 
\begin{eqnarray}
x_*(s) = \left( \frac{ s }{c \eta } \right)^{\frac{1}{\eta-1}}
\label{xetoile}
\end{eqnarray}
One obtains
\begin{eqnarray}
{\hat  \Pi_b}(q=is) 
 \opsimeq_{s \to + \infty}
   \sqrt{ \frac{2 \pi}{ c \eta (\eta-1) x^{\eta-2}_*(s) }}
e^{ c (\eta-1)x^{\eta}_*(s)  + \Phi(x_*(s))  } 
\label{fbsaddle}
\end{eqnarray}

\subsubsection{ Use of the convolution equation} 

The convolution equation is simple in Fourier ( Eq. \ref{convolfourier})
\begin{eqnarray}
{\hat  G_b}(q=is) = ( {\hat  \Pi_b}(q=is)  )^2 \opsimeq_{s \to + \infty} 
=   \frac{2 \pi}{ c \eta (\eta-1) x^{\eta-2}_*(s) }
e^{ 2 c (\eta-1)x^{\eta}_*(s)  + 2 \Phi(x_*(s))  }
\label{gbsaddle}
\end{eqnarray}

\subsubsection{ Use of the recursion equation} 

We now consider
 the recursion of Eq. \ref{recursionscal} in the tail
$ u \to - \infty$.  Using the normalization 
\begin{eqnarray}
\int_{-\infty}^{+\infty} dx \Pi_b(x) =
1 = \int_{-\infty}^{+\infty} dx G_b(x) 
\label{normapdf}
\end{eqnarray}
we may rewrite it as
\begin{eqnarray}
1-\int_{-\infty}^{\frac{u}{\lambda}} dx \Pi_b(x) 
= \left[ 1- \int_{-\infty}^{u} dx G_b(x) \right]^b
\label{recursiontail}
\end{eqnarray}
At leading order in the tail, one has
\begin{eqnarray}
\int_{-\infty}^{\frac{u}{\lambda}} dx \Pi_b(x) 
= b \int_{-\infty}^{u} dx G_b(x) +... 
\label{tailneg}
\end{eqnarray}
The identification 
\begin{eqnarray}
\frac{1}{\lambda}  \Pi_b \left(\frac{u}{\lambda} \right)
 \opsimeq_{u \to -\infty} 
b  G_b(u) 
\label{fbgbtail}
\end{eqnarray}
becomes in Fourier
\begin{eqnarray}
{\hat  \Pi_b} ( q=i \lambda s) \opsimeq_{s \to +\infty} 
b {\hat  G_b}(q=is)
 \label{fbgbtailfourier}
\end{eqnarray}

\subsubsection{ Identification} 

Rewriting Eq. \ref{fbgbtailfourier}
using Eq. \ref{fbsaddle}  and Eq. \ref{gbsaddle}
yields the following constraints.
The identification of the leading term in the exponential yields
\begin{eqnarray}
 \lambda^{\frac{\eta}{\eta-1}}  =   2  
\label{leading}
\end{eqnarray}
and with the notation $\lambda \equiv 2^{\omega}$ (Eq.  \ref{deltan} ) 
this gives
\begin{eqnarray}
 \eta= \frac{1}{1-\omega}
\label{leadingsuite}
\end{eqnarray}
The identification of subleading terms is compatible with
the power-law form
\begin{eqnarray}
e^{ \Phi(x) } \simeq  A x^{- \nu}
\label{powerlawform}
\end{eqnarray}
with the parameters
\begin{eqnarray}
 \nu && = \frac{2 -\eta}{2} \nonumber \\
 A && = \frac{1}{b} \sqrt{ \frac{c \eta (\eta-1)}{ 2 \pi} }
\label{hA}
\end{eqnarray}

In Fourier, this corresponds to pure exponential forms
\begin{eqnarray}
{\hat  \Pi_b}(q=is) 
&& \opsimeq_{s \to + \infty}   \frac{1}{ b }
e^{ c (\eta-1) \left( \frac{ s }{ c \eta} \right)^{\frac{\eta}{\eta-1}}   } \nonumber \\
{\hat  G_b}(q=is) 
&& = ({\hat  \Pi_b}(q=is) )^2 \opsimeq_{s \to + \infty}
  \frac{1}{ b^2 }
e^{ 2 c (\eta-1) \left( \frac{ s }{ c \eta} \right)^{\frac{\eta}{\eta-1}}   }
\label{resleftfourier}
\end{eqnarray}

\subsection{ Study of the right-tail form $\Pi_b(x \to +\infty)$ }

We write the right-tail of $\Pi_b(x)$ as
\begin{eqnarray}
 \Pi_b(x) \opsimeq_{x \to + \infty}
  e^{- d   x^{\eta'} + \Psi(  x ) }
\label{fbrighttail}
\end{eqnarray}
where the function $\Psi( x )$ is subleading 
with respect to the first term of order $ x^{\eta'}$.
This right tail will dominate the Fourier transform for $q=-i s$
with real $s \to +\infty$
\begin{eqnarray}
{\hat \Pi_b}(q=-is) && \equiv \int_{-\infty}^{+\infty} dx e^{  s x } \Pi_b(x)
\opsimeq_{s \to + \infty} \int dx e^{  s x 
- d x^{\eta'}+ \Psi(x)  } 
\label{pnsr}
\end{eqnarray}
Since $\Psi(x)$ is subleading, 
we perform a saddle point calculation in the two first terms,
yielding the saddle value 
\begin{eqnarray}
x_+(s) = \left( \frac{ s }{ d \eta' } \right)^{\frac{1}{\eta'-1}}
\label{xetoiler}
\end{eqnarray}
One obtains
\begin{eqnarray}
{\hat  \Pi_b}(q=-is) 
 \opsimeq_{s \to +\infty}
=  \sqrt{ \frac{2 \pi}{d \eta' (\eta'-1) x^{\eta'-2}_+(s)  }}
e^{ d (\eta'-1)x^{\eta'}_+(s) + \Psi(x_+(s))  }
\label{fbsaddler}
\end{eqnarray}

\subsubsection{ Use of the convolution equation} 

The convolution equation is simple in Fourier ( Eq. \ref{convolfourier})
\begin{eqnarray}
{\hat  G_b}(q=-is) = ({\hat  \Pi_b}(q=-is))^2 \opsimeq_{s \to +\infty} 
\frac{2 \pi}{d \eta' (\eta'-1) x^{\eta'-2}_+(s) }
 e^{ 2 d (\eta'-1)x^{\eta'}_+(s) + 2 \Psi(x_+(s))  } 
\label{gbsaddler}
\end{eqnarray}
The asymptotic behavior of $G_b$ for $x \to \infty$ is then
\begin{eqnarray}
 G_b(x) \opsimeq_{ x \to +\infty}  e^{- D  x^{\eta'} + \rho(  x ) }
\label{gbsaddlerealspace}
\end{eqnarray}
with the correspondence
\begin{eqnarray}
 D && = d \ \ 2^{ 1- \eta' } \nonumber \nonumber \\
e^{ \rho(2 x) - 2 \Psi \left( x \right)} && = 
 \sqrt{ \frac{  \pi }{ d \eta' (\eta'-1) x^{\eta'-2} }}
\label{gbpara}
\end{eqnarray}
This suggests the power-law forms
\begin{eqnarray}
e^{ \Psi(x) } && \simeq  B x^{- \kappa} \nonumber \\
e^{ \rho(x) } && \simeq  {\cal B} x^{- \sigma }
\label{gbparabis}
\end{eqnarray}
with the following relations between exponents and amplitudes
\begin{eqnarray}
\sigma && = 2 \kappa+\frac{\eta'-2}{2} \nonumber \nonumber \\
{\cal B}  && =  B^2 2^{ \sigma }
 \sqrt{ \frac{  \pi }{ d \eta' (\eta'-1)  }}
\label{sigmacalb}
\end{eqnarray}

\subsubsection{ Use of the recursion equation} 

We now consider
 the recursion of Eq. \ref{recursionscal} in the tail
$ u \to + \infty$.
We may use Eq. \ref{fbrighttail}
and use a saddle-point calculation at the left boundary $x_{left}=u/\lambda$
yielding 
\begin{eqnarray}
\int_{\frac{u}{\lambda}}^{+\infty} dx \Pi_b(x) 
\opsimeq_{u \to \infty} \frac{1}{ d \eta' 
\left( \frac{u}{\lambda} \right)^{\eta'-1}} \ \ 
e^{- d   \left( \frac{u}{\lambda} \right)^{\eta'}
 + \Psi \left( \frac{u}{\lambda} \right) }
\end{eqnarray}
Similarly from Eq. \ref{gbsaddlerealspace}
\begin{eqnarray}
\int_{u}^{+\infty} dx G_b(x) 
\opsimeq_{u \to \infty} \frac{1}{ D \eta' u^{\eta'-1}} \ \ 
e^{-  D u^{\eta'}
 + \rho (u) }
\end{eqnarray}

So Eq. \ref{recursionscal} becomes
\begin{eqnarray}
 \frac{1}{ d \eta' 
\left( \frac{u}{\lambda} \right)^{\eta'-1}} \ \ 
e^{- d   \left( \frac{u}{\lambda} \right)^{\eta'}
 + \Psi \left( \frac{u}{\lambda} \right) }
= \left[  \frac{1}{ D \eta' u^{\eta'-1}} \right]^b  e^{- b D u^{\eta'}
 + b \rho (u) }
\label{recursionscalrasymp}
\end{eqnarray}

\subsubsection{ Identification} 

The identification of the leading term gives using Eq. \ref{gbpara}
\begin{eqnarray}
\lambda^{\eta'} = \frac{d}{ b D} = \frac{ 2^{\eta'-1} }{ b }
\label{leadingright}
\end{eqnarray}
Comparison with Eq. \ref{leading} yields
that the two exponents $\eta$ and $\eta'$ are related via
\begin{eqnarray}
2^{\frac{\eta'}{\eta}} = (2b)
\label{etamu}
\end{eqnarray}

The subleading terms yield
\begin{eqnarray}
e^{ b \rho (u) - \Psi \left( \frac{u}{\lambda} \right) }
=   \frac{ ( D \eta' u^{\eta'-1} )^b }
{ d \eta'  \left( \frac{u}{\lambda} \right)^{\eta'-1}}  
= ( 2 d \eta' \  u^{\eta'-1} )^{b-1} (2 b)^{\frac{1-\eta'}{\eta'}}
\label{subleadingright}
\end{eqnarray}

Using the power-law forms of
Eqs  \ref{gbparabis}, one obtains via identification
\begin{eqnarray}
\kappa -b \sigma && = (b-1) (\eta'-1) \nonumber \\
\frac{ ( {\cal B})^b }{ B \lambda^{\kappa} } &&= ( 2 d \eta' )^{b-1} (2 b)^{\frac{1-\eta'}{\eta'}}
\end{eqnarray}

Consistency with Eq. \ref{sigmacalb} yields
\begin{eqnarray}
\kappa &&  =  \frac{1}{2b-1} \left[ -\frac{b}{2} (\eta'-2) - (b-1)(\eta'-1)  \right] \nonumber \\
\sigma &&  =  \frac{1}{2b-1} \left[ -\frac{1}{2} (\eta'-2) - 2(b-1)(\eta'-1)  \right]
\end{eqnarray}

\section{Reminder on Zhang argument
 for the left tail of the directed polymer }  

\label{zhangargument}

Let us now recall Zhang's argument \cite{Hal_Zha} that allows to determine
the exponent $\eta$ of the left tail of the free energy distribution
of the directed polymer
\begin{eqnarray}
P_L(F \to -\infty) \sim e^{- \left( \frac{ \vert F \vert}{L^{\omega}} \right)^{\eta} } 
\label{taileta}
\end{eqnarray}
Positive moments of the partition function can be evaluated by the saddle-point
method, with a saddle value $F^*$ lying in the negative tail (\ref{taileta})
\begin{eqnarray}
\overline{ Z_L^k} = \int dF P_L(F) e^{ - k \beta F_L}  \sim \int dF 
e^{- \left( \frac{ \vert F \vert}{L^{\omega}} \right)^{\eta} } e^{ - k \beta F_L} 
\sim e^{ c(k) L^{ \frac{ \omega \eta}{\eta-1}   } }
\label{saddle}
\end{eqnarray}
Since these moments of the partition function 
have to diverge exponentially in $L$,
the exponent $\eta$ of the tail (\ref{taileta})
 reads in terms of the droplet exponent
\begin{eqnarray}
\eta=\frac{1}{1-\omega}
\label{etazhang} 
\end{eqnarray}

This argument can be extended to the free-energy distribution of other
random systems. However, as recalled in the introduction, 
the distribution of the free-energy over the samples
is simply Gaussian with $\eta=2$ and $\omega_f=d/2$
for spin models in any finite dimension $d$.
The distribution of the ground state energy is non Gaussian
for the mean-field Sherrington-Kirkpatrick model of spin-glasses
and we refer to \cite{DPlefttail}
for a discussion of the corresponding tail exponent $\eta$.


\begin{thebibliography}{99}

\bibitem{Gum_Gal}
E.J. Gumbel, `` Statistics of extreme'' (Columbia University Press, NY 1958);
J. Galambos, `` The asymptotic theory of extreme order statistics'' 
( Krieger , Malabar, FL 1987).

\bibitem{Bou_Krz_Mar} J.-P. Bouchaud, F. Krzakala and O.C. Martin,
Phys. Rev. {\bf B68}, 224404 (2003).

\bibitem{Fis_Hus_SG}
D.S. Fisher and D.A. Huse, Phys. Rev. {\bf B38}, 386 (1988).

\bibitem{Fis_Hus}
D.S. Fisher and D.A. Huse,  Phys. Rev. {\bf B43}, 10728  (1991).


\bibitem{We_Ai}
J. Wehr and M. Aizenman, J. Stat. Phys. 60 (1990) 287.


\bibitem{Hal_Zha}
T. Halpin-Healy and Y.-C. Zhang, Phys. Repts., {\bf 254}, 215 (1995).



\bibitem{DPexcita}
C. Monthus and T. Garel, Phys. Rev. E 73 , 056106 (2006).




\bibitem{Hus_Hen_Fis}
D. A. Huse, C. L. Henley, and D. S. Fisher, 
Phys. Rev. Lett. 55, 2924 (1985).

\bibitem{Kar}
M. Kardar, Nucl. Phys. B {\bf 290} 582 (1987).

\bibitem{Joh}
K. Johansson, Comm. Math. Phys. 209 (2000) 437.
 
\bibitem{Pra_Spo}
 M. Prahofer and H. Spohn,  Physica A 279, 342 (2000) ; 
M. Prahofer and H. Spohn, Phys. Rev. Lett. 84, 4882   (2000) ; 
    M. Prahofer and H. Spohn, J. Stat. Phys. 108, 1071 (2002)  ; 
  M. Prahofer and H. Spohn, cond-mat/0212519.

\bibitem{prae}
M. Pr\"ahoher and H. Spohn, http://www-m5.ma.tum.de/KPZ/.




\bibitem{realspaceRG}
Th. Niemeijer, J.M.J. van Leeuwen, ''Renormalization theories for Ising
spin systems'' in Domb and Green Eds, ''Phase Transitions and Critical
 Phenomena'' (1976); T.W. Burkhardt and J.M.J. van Leeuwen, 
``Real-space renormalizations'', Topics in current Physics,
 Vol. 30, Spinger, Berlin (1982);
B. Hu, Phys. Rep. 91, 233 (1982).

\bibitem{MKRG}
A.A. Migdal, Sov. Phys. JETP 42, 743 (1976) ; 
L.P. Kadanoff, Ann. Phys. 100, 359 (1976).

\bibitem{berker}
A.N. Berker and S. Ostlund, J. Phys. C 12, 4961 (1979).

\bibitem{hierarchical}
M. Kaufman and R. B. Griffiths, Phys. Rev. B 24, 496 - 498 (1981);
    R. B. Griffiths and M. Kaufman, Phys. Rev. B 26, 5022  (1982).


\bibitem{diluted}
C. Jayaprakash, E. K. Riedel and M. Wortis, 
Phys. Rev. B 18, 2244 (1978)


\bibitem{Kin_Dom}
W. Kinzel and E. Domany, Phys. Rev. B 23, 3421 (1981).

\bibitem{Der_Potts}
B. Derrida and E. Gardner, J. Phys. A 17, 3223 (1984);
B. Derrida, Les Houches (1984).

\bibitem{andelman}
D. Andelman and A.N. Berker, Phys. Rev. B 29, 2630 (1984).



\bibitem{young}
A. P. Young and R. B. Stinchcombe,
J. Phys. C 9 (1976) 4419 ; 
B. W. Southern and A. P. Young
J. Phys. C 10 ( 1977) 2179.

\bibitem{mckay}
S.R. McKay, A.N. Berker and S. Kirkpatrick,
Phys. Rev. Lett. 48 (1982) 767;
E. J. Hartford, J. Appl. Phys. 70, 6068 (1991).

\bibitem{Gardnersg}
E. Gardner, J. Physique 45, 115 (1984).


\bibitem{bray_moore}
A.J. Bray and M. A. Moore, J. Phys. C 17 (1984) L463;
J.R. Banavar and A.J. Bray, Phys. Rev. B 35, 8888 (1987);
M. A. Moore, H. Bokil, B. Drossel
    Phys. Rev. Lett. 81 (1998) 4252.



\bibitem{nifle_hilhorst}
M. Nifle and H.J. Hilhorst, Phys. Rev. Lett. 68 (1992) 2992 ;
M. Ney-Nifle and H.J. Hilhorst, Physica A 193 (1993) 48 ;
M.J. Thill and H.J. Hilhorst, J. Phys. I France 6, 67 (1996)




\bibitem{Der_Gri}
B. Derrida and R.B. Griffiths, Eur.Phys. Lett. 8 , 111 (1989).

\bibitem{Coo_Der}
J. Cook and B. Derrida, J. Stat. Phys. 57, 89 (1989).

\bibitem{Tim}
T. Halpin-Healy, Phys. Rev. Lett. 63, 917 (1989);
Phys. Rev. A , 42 , 711 (1990).

\bibitem{roux}
S. Roux, A. Hansen, L R da Silva, LS Lucena and RB Pandey,
J. Stat. Phys. 65, 183 (1991).

\bibitem{kardar}
L. Balents and M. Kardar, J. Stat. Phys. 67, 1 (1992);
E. Medina and M. Kardar, J. Stat. Phys. 71, 967 (1993).

\bibitem{cao}
M.S. Cao, J. Stat. Phys. 71, 51 (1993).

\bibitem{tang}
LH Tang J Stat Phys 77, 581 (1994).

\bibitem{Muk_Bha}
    S. Mukherji and S. M. Bhattacharjee,
Phys. Rev. E 52, 1930 (1995).

\bibitem{Bou_Sil}
 R. A. da Silveira and J. P. Bouchaud,
Phys. Rev. Lett. 93, 015901 (2004)





\bibitem{Tan_For_Wol}
L.H. Tang, B.M. Forrest and D.E. Wolf, Phys. Rev. A 45 (1992) 7162.

\bibitem{Ala_etal}
T. Ala-Nissila, T. Hjelt, J.M. Kosterlitz and V. Venalainen, J. Stat. Phys. 72
(1993) 207.

\bibitem{KimetAla}
T. Ala-Nissila,  Phys. Rev. Lett. 80 (1998) 887 ;
J.M. Kim, Phys. Rev. Lett. 80 (1998) 888.

\bibitem{Mar_etal}
E. Marinari, A. Pagnani and G. Parisi, J Phys. A 33 (2000) 8181 ;
E. Marinari, A. Pagnani and G. Parisi and Z. Racz, 
Phys. Rev. E 65 (2002) 026136.

\bibitem{Las_Kin}
M. Lassig and H. Kinzelbach, Phys. Rev. Lett. 78 (1997) 903.

\bibitem{Col_Moo}
F. Colaiori and M. A. Moore, Phys. Rev. Lett. 86 (2001) 3946.

\bibitem{LeDou_Wie}
P. Le Doussal and K. Wiese, Phys. Rev. E 72 (2005) 035101.


\bibitem{diamondpolymer}
C. Monthus and T. Garel, arXiv:0710.0735.


\bibitem{DPlefttail}
C. Monthus and T. Garel,
Phys. Rev. E 74, 051109 (2006).


\bibitem{Hus_Hen}
D. A. Huse, C. L. Henley, Phys. Rev. Lett. 54, 2708 (1985).




\end{thebibliography}
\end{document}